\documentclass[twoside,10pt]{article}
\usepackage{CJK}
\usepackage{indentfirst}
\usepackage{bm}
\usepackage[]{hyperref}
\usepackage[numbers,sort&compress]{natbib}
\usepackage{tikz}
\usepackage{float}
\usepackage{color}
\usepackage{graphicx}
\usepackage{subfig}
\usepackage{amsmath}
\usepackage{mathrsfs}
\usepackage{amssymb}
\usepackage{amsthm}
\newcommand{\RNum}[1]{\uppercase\expandafter{\romannumeral #1\relax}}

\makeatletter
\@addtoreset{equation}{section}
\makeatother
\begin{document}
\begin{CJK*}{GBK}{song}
\allowdisplaybreaks


\begin{center}
\LARGE\bf  Spectral analysis and long-time asymptotics of complex mKdV equation
\end{center}
\footnotetext{\hspace*{-.45cm}\footnotesize $^\dag$Corresponding author: Y.F. Zhang. E-mail: zhangyfcumt@163.com }
\begin{center}
\ \\Hong-Yi Zhang, Yu-Feng Zhang
\end{center}
\begin{center}
\begin{small} \sl
{School of Mathematics, China University of Mining and Technology, Xuzhou, Jiangsu, 221116, People's Republic of
China.\\}
\end{small}
\end{center}
\vspace*{2mm}
\begin{center}
\begin{minipage}{15.5cm}
\parindent 20pt\footnotesize


 \noindent {\bfseries
Abstract} In this paper, we obtain the long-time asymptotics of complex mKdV equation via Defit-Zhou method (Nonlinear steepest descent method). The Cauchy problem of complex mKdV equation is transformed into the corresponding Riemann-Hilbert problem on the basis of the Lax pair and the scattering matrix. After that Riemann-Hilbert problems are converted through a decomposition of the matrix-valued spectral function and factorizations of the jump matrix for Riemann-Hilbert problem. Finally, by solving the last model problem, the long-time asymptotics of complex mKdV equation are derived.

\end{minipage}
\end{center}
\begin{center}
\begin{minipage}{15.5cm}
\begin{minipage}[t]{2.3cm}{\bf Keywords}\end{minipage}
\begin{minipage}[t]{13.1cm}
Riemann-Hilbert problem; Complex mKdV equation; Nonlinear steepest descent method; Long-time asymptotics
\end{minipage}\par\vglue8pt
\end{minipage}
\end{center}

\section{Introduction}  
The research of nonlinear partial differential equations (NLPDEs) has played an important role in the development of science and technology. Nowadays, NLPDEs can be used to explain some complex physical phenomena, including mathematics, plasma physics, aerodynamics, fluid mechanics, atmospheric oceans, etc [1-8]. Hirota bilinear method[9,10], Darboux transformation[11,12], the inverse scattering transformation[13] and so on are effective method to solve NLPDEs. Especially, the inverse scattering transformation is the first method found to be used to obtain the exact solution of the soliton equation, and it is called an effective method for analyzing the initial value of the integrable evolution equation on a straight line[14-16]. In 1993, Defit and Zhou proposed the famous nonlinear steepest descent method to analyze the long-time asymptotic behavior of integrable evolution equations[17]. And Defit and Zhou analyzed the long-time asymptotic behavior of the solution to the initial value problem of the famous mKdV equation and Schr\"{o}dinger equation[17,18]. Subsequently, the nonlinear steepest descent method was extended to the study of integrable evolution equations. In recent years, Fan and Geng researched some long-time asymptotic behavior of integrable evolution equations based on the method[19-24].

In this paper, we study the equation derived from the Lax pair given by Yishen Li[25]. The Lax pair is
\begin{equation}
\begin{aligned}
&{\psi_x}=-iz{\sigma_3}\psi+P\psi,\\
&{\psi_t}=(\vartheta z^{3}+\iota z^{2}+\kappa z+\varsigma){\sigma_3}\psi+Q\psi,
\end{aligned}
\end{equation}
where $\psi(z;x,t)$ is a $2\times 2$ matrix, ${\sigma_3}$=diag$(1,-1)$, and
\begin{equation}
\begin{aligned}
&~~~~~~~~~~~~~~~~~~~~~~~~~~~~~~~~~~~~~~~~~~~~~~~~~P=\left(
\begin{array}{cc}
0 & u\\
v & 0\\
 \end{array}
 \right) ,\\
&Q=i\vartheta{z^{2}}P -iz\left(
\begin{array}{cc}
\frac{i\vartheta}{2}uv & -\frac{i\vartheta}{2}{u_x}-\iota u \\
\frac{i\vartheta}{2}{v_x}-\iota v & -\frac{i\vartheta}{2}uv \\
 \end{array}
 \right)
 -\left(
\begin{array}{cc}
\frac{i\vartheta}{4}(u{v_x}-v{u_x})-\frac{\iota}{2}uv & -\frac{i\vartheta}{4}(-{u_{xx}}+2u^{2}v)+\frac{\iota}{2}{u_x}-i\kappa u \\
\frac{-i\vartheta}{4}(-{v_{xx}}+2uv^{2})-\frac{\iota}{2}{v_x}-i\kappa v & -\frac{i\vartheta}{4}(u{v_x}-v{u_x})+\frac{\iota}{2}uv\\
 \end{array}
 \right).\\
 \end{aligned}
\end{equation}

The Lax pair (1.1) derives the following equation:
\begin{equation}
\begin{cases}
{u_t}=-\frac{i\vartheta}{4}({u_{xxx}}-6uv{u_x})-\frac{\iota}{2}({u_{xx}}-2u^{2}v)+i\kappa {u_x}+2\varsigma u,\\
{v_t}=-\frac{i\vartheta}{4}({v_{xxx}}-6uv{v_x})+\frac{\iota}{2}({v_{xx}}-2v^{2}u)+i\kappa {v_x}-2\varsigma v.\\
\end{cases}
\end{equation}
(I)Take $\vartheta=-4i$, $\iota=\kappa=\varsigma=0$, $v=-1$. Eq.(1.3) reduces to KdV equation
\begin{equation}
{u_t}+6u{u_x}+{u_{xxx}}=0.
\end{equation}
(II)Take $\vartheta=-4i$, $\iota=\kappa=\varsigma=0$, $v=-u$. Eq.(1.3) reduces to the mKdV equation
\begin{equation}
{u_t}+6u^{2}{u_x}+{u_{xxx}}=0.
\end{equation}
(III)Take $\iota=-2i$, $\vartheta=\kappa=\varsigma=0$, $v=\mp\overline{u}$. Eq.(1.3) reduces to nonlinear Schr\"{o}dinger equation
\begin{equation}
i{u_t}+{u_{xx}}\pm2u^{2}\overline{u}=0,
\end{equation}
where superscript bar denotes complex conjugate.\\
(IV)Take $\iota=-2$, $\vartheta=\kappa=\varsigma=0$, ${q_x}=uv={\left(\frac{{u_x}}{u}\right)}{_x}$. Eq.(1.3) reduces to Burger equation
\begin{equation}
{q_t}=2q{q_x}-{q_{xx}}.
\end{equation}
Especially, take $\vartheta=-i\alpha$ ($\alpha>0$), $\iota=\kappa=\varsigma=0$ and $v=\overline{u}$. Eq.(1.3) reduces to complex mKdV equation
\begin{equation}
{u_t}=\frac{\alpha}{4}(-{u_{xxx}}+2{(|u|^{2}u)}{_x}),
\end{equation}
where $u(x,t)$ is complex-valued function of variate $(x,t)$.

In this paper, we use Defit-Zhou nonlinear steepest descent method to study long-time asymptotics of Eq.(1.8) with Schwartz decaying initial data\\
$~~~~~~~~~~~~~~~~~~~~~~~~~~~~~~~~~~~~~~~~~~~~~~~~~~~~~~u(t=0,x)={u_0}(x)\in \mathcal{S}(\mathbb{R}).$\\
We know that long-time asymptotics of Eq.(1.8) has not been researched, the main result is presented in the following theorem.\\
\textbf{Theorem 1.1} $u(x,t)$ be the solution for the Cauchy problem of complex mKdV equation (1.8) with ${u_0} \in \mathcal{S}(\mathbb{R})$, where $\mathcal{S}(\mathbb{R})$ represents Schwartz space:
$\mathcal{S}(\mathbb{R})={f\in C^{\infty}(R),~~ {\|f\|}{_{\alpha,\beta}}=\sup |x^{\alpha}\partial^{\beta}f(x)|<\infty,~~\alpha,\beta\in {Z_+}}.$
For $x<0$ and $\frac{x}{t} \leq C$, the leading asymptotics of $u(x, t)$ has the form:
\begin{equation}
u(x,t)=\frac{\upsilon e^{\frac{\pi\upsilon}{2}}}{\sqrt{6\alpha t {z_0}\pi}}\left((\delta_{B}^{0})^{2}e^{-\frac{3\pi i}{4}}\Gamma(i\upsilon)\overline{r({z_0})}-(\delta_{B}^{0})^{-2}e^{\frac{3\pi i}{4}}\Gamma(-i\upsilon)r({z_0})\right)+o\left(c({z_0})t^{-1}\log t\right)
\end{equation}
where $C$ is a constant, $\Gamma$ is the Gamma function, $c$ is a rapidly decreasing function, and\\
$~~~~~~~~~~~~~~~~~~~~~~~~~~~~~~~~~~~~~~~~~~\delta_{B}^{0}=\left(12\alpha t({z_0})^{3}\right)^{-\frac{i\upsilon}{2}}e^{2i\alpha t{z_0}^{3}}e^{\chi({z_0})}$,\\
$~~~~~~~~~~~~~~~~~~~~~~~~~~~~~~~~~~~~~~~~~~{z_0}=\sqrt{-\frac{x}{3\alpha t}}$,~~$\upsilon=-\frac{1}{2\pi}\log\left(1-|r({z_0})|^{2}\right)$,\\
$~~~~~~~~~~~~~~~~~~~~~~~~~~~~~~~~~~~~~~~~~~\chi({z_0})=\frac{1}{2\pi i}\int_{-{z_0}}^{{z_0}} \log\left(\frac{1-|r(\xi)|^{2}}{1-|r({z_0})|^{2}}\right)\frac{d\xi}{\xi-z}$.\\

The outline of this paper is as follows. In section 2, we analyze eigenfunction and spectral function of Eq.(1.8) to construct a original Riemann-Hilbert problem. In section 3, by deforming the jump matrix of the original Riemann-Hilbert problem and extending region, the original Riemann-Hilbert problem is transformed into a model Riemann-Hilbert problem. Then the solution of the model Riemann-Hilbert problem can be expressed by the solution of Weber equation. Finally, we obtain the long-time asymptotics of the Cauchy problem for complex mKdV equation.

\section{The Riemann-Hilbert Problems}  
In this section, we first make use of the Lax pair of complex mKdV equation to construct the matrix Jost solutions. Then the Cauchy problem of complex mKdV equation turns into Riemann-Hilbert problem.

By the transformation
\begin{equation}
\mu(z;x,t)=\psi(z;x,t)e^{izx{\sigma_3}+i\alpha z^{3}t{\sigma_3}},
\end{equation}
 $\mu$ satisfies the following Lax pair
\begin{equation}
\begin{aligned}
&{\mu_x}=iz[\mu,{\sigma_3}]+P\mu,\\
&{\mu_t}=i\alpha z^{3}[\mu,{\sigma_3}]+Q\mu,\\
\end{aligned}
\end{equation}
where $e^{{\sigma_3}}$=diag$(e,e^{-1})$, $[\mu,{\sigma_3}]=\mu{\sigma_3}-{\sigma_3}\mu $.
Lax pair (2.2) can be written as the following full derivative form
\begin{equation}
d\left(e^{(izx+i\alpha z^{3}t)\hat{\sigma}_{3}}\mu\right)=e^{(izx+i\alpha z^{3}t)\hat{\sigma}_{3}}[(Pdx+Qdt)\mu],\\
\end{equation}
where $e^{{\hat{\sigma}}{_3}}u=e^{{\sigma_3}}ue^{{-\sigma_3}}$.
To analyze the eigenfunction $\mu(z;x,t)$, we choose two special integral paths
\begin{equation}
(-\infty,t)\to (x,t),~~and~~(\infty,t)\to (x,t),
\end{equation}
and acquire two eigenfunctions of Lax pair (2.2)
\begin{equation}
\begin{aligned}
&{\mu_1}(z;x,t)=I+\int_{-\infty}^{x}e^{-iz(x-\xi)\hat{\sigma}_{3}}P(z;\xi,t){\mu_1}(z;\xi,t)d\xi,\\
&{\mu_2}(z;x,t)=I-\int_{x}^{\infty}e^{-iz(x-\xi)\hat{\sigma}_{3}}P(z;\xi,t){\mu_2}(z;\xi,t)d\xi.\\
\end{aligned}
\end{equation}
${\mu_1}(z;x,t)$ and ${\mu_2}(z;x,t)$ have the following asymptotics
\begin{equation}
\begin{aligned}
&{\mu_1}(z;x,t),{\mu_2}(z;x,t)\to ~ I,~~x \to \pm\infty,\\
&{\mu_1}(z;x,t),{\mu_2}(z;x,t)\to ~ I,~~z \to \infty.\\
\end{aligned}
\end{equation}
There are two first order linear homogeneous equations in Lax pair (1.1),
while ${\psi_1}(z;x,t)={\mu_1}(z;x,t)e^{-i\theta(z)\hat{\sigma}_{3}}$ and ${\psi_2}(z;x,t)={\mu_2}(z;x,t)e^{-i\theta(z)\hat{\sigma}_{3}}$
are solutions of Lax pair (1.1), then ${\psi_1}(z;x,t)$ and ${\psi_2}(z;x,t)$ are linearly correlated. Thus
\begin{equation}
{\mu_1}(z;x,t)={\mu_2}(z;x,t)e^{-i\theta(z)\hat{\sigma}_{3}}S(z),
 \end{equation}
where \\
$~~~~~~~~~~~~~~~~~~~~~~~~~~~~~\theta(z)=zx+\alpha z^{3}t$,
$~~S(z)=\left(
\begin{array}{cc}
{s_{11}}(z) & {s_{12}}(z)\\
{s_{21}}(z) & {s_{22}}(z)\\
 \end{array}
 \right)$.\\
 $S(z)$ is irrelevant to $x$ and $t$, it is called as the spectral matrix function. Nextly, we study the analyticity of ${\mu_1}(z;x,t)$, ${\mu_2}(z;x,t)$ and $S(z)$. For integral equation (2.5), we have
 \begin{equation}
e^{-iz(x-\xi)\hat{\sigma}_{3}}P(z;\xi,t)=\left(
\begin{array}{cc}
0 &  ue^{-2iz(x-\xi)}\\
\overline{u}e^{2iz(x-\xi)} & 0\\
 \end{array}
 \right), ~~\text{as}~~x> \xi,
\end{equation}
where\\
$~~~~~~~~~~~~~~~~~~~~~~~~~~~e^{2iz(x-\xi)}=e^{2i(x-\xi)Rez}e^{-2(x-\xi)Imz}$,~~~$e^{-2iz(x-\xi)}=e^{-2i(x-\xi)Rez}e^{2(x-\xi)Imz}$.\\
After direct calculation, we note that the first column of ${\mu_1}(z;x,t)$ is analytical in the upper half plane ${\mathbb{C}_+}$, the second column of ${\mu_1}(z;x,t)$ is analytical in the lower half plane ${\mathbb{C}_-}$, where ${\mu_1}(z;x,t)$ can be written as
 \begin{equation}
{\mu_1}=\left(
\begin{array}{cc}
{\mu}_1^{(11)} &  {\mu}_1^{(12)}\\
{\mu}_1^{(21)} & {\mu}_1^{(22)}\\
 \end{array}
 \right)=({\mu}_1^{+},{\mu}_1^{-}).
\end{equation}
Similarly, the first column of ${\mu_2}(z;x,t)$ is analytical in the lower half plane ${\mathbb{C}_-}$ and the second column of ${\mu_2}(z;x,t)$ is analytical in the upper half plane ${\mathbb{C}_+}$, where ${\mu_2}(z;x,t)$ can be written as
\begin{equation}
{\mu_2}=\left(
\begin{array}{cc}
{\mu}_2^{(11)} &  {\mu}_2^{(12)}\\
{\mu}_2^{(21)} & {\mu}_2^{(22)}\\
 \end{array}
 \right)=({\mu}_2^{-},{\mu}_2^{+}).
\end{equation}
By using Abel formula, we get
 \begin{equation}
\begin{aligned}
&{\left(\det{\psi_j}\right)_x}=\operatorname{tr}\left(P-iz{\sigma_3}\right)\det{\psi_j}=0,\\
&{\left(\det{\psi_j}\right)_t}=\operatorname{tr} \left(Q-i\alpha z^{3}{\sigma_3}\right)\det{\psi_j}=0.\\
\end{aligned}
\end{equation}
Through transformation (2.1),
\begin{equation}
\det{\mu_j}=\det{\psi_j}\det\left(e^{i\theta(z){\sigma_3}}\right)=\det{\psi_j}.
\end{equation}
Therefore, we obtain
\begin{equation}
{\left(\det{\mu_j}\right)_x}={\left(\det{\mu_j}\right)_t}=0,
\end{equation}
which imply that $\det{\mu_j}$ is independent of $x$ and $t$, and ${\mu_j}\rightarrow I$, $|x|\rightarrow \infty$. So we have
\begin{equation}
\det{\mu_j}=\lim \limits_{|x|\to \infty}\det{\mu_j}=\det\left(\lim \limits_{|x|\to \infty}{\mu_j}\right)=1.
\end{equation}
Taking determinants on both sides of Eq.(2.7), yields
\begin{equation}
\det S(z)=1.
\end{equation}
From Eq.(2.14), we know that ${\mu_1}(z;x,t)$, ${\mu_2}(z;x,t)$ are invertible matrices. Based on the analyticity of the column vector for ${\mu_1}(z;x,t)$ and ${\mu_2}(z;x,t)$, it can be inferred that the first and second row of ${\mu}_1^{-1}(z;x,t)$ are analytical in the lower half plane ${\mathbb{C}_-}$ and the upper half plane ${\mathbb{C}_+}$ respectively. The first and second row of ${\mu}_2^{-1}(z;x,t)$ are analytical in the upper half plane ${\mathbb{C}_+}$ and the lower half plane ${\mathbb{C}_-}$ respectively. That is
\begin{equation}
{\mu}_1^{-1}=\left(
\begin{array}{cc}
{\mu}_1^{(22)} &  -{\mu}_1^{(12)}\\
-{\mu}_1^{(21)} & {\mu}_1^{(11)}\\
 \end{array}
 \right)=\binom{{\hat{\mu}}_1^{-}}{{\hat{\mu}}_1^{+}},
\end{equation}

\begin{equation}
{\mu}_2^{-1}=\left(
\begin{array}{cc}
{\mu}_2^{(22)} &  -{\mu}_2^{(12)}\\
-{\mu}_2^{(21)} & {\mu}_2^{(11)}\\
 \end{array}
 \right)=\binom{{\hat{\mu}}_2^{+}}{{\hat{\mu}}_2^{-}}.
\end{equation}
Through Eq.(2.7), Eq.(2.16) and Eq.(2.17), we get
\begin{equation}
e^{-i\theta(z)\hat{\sigma}_{3}}S(z)={\mu}_2^{-1}{\mu_1}=\binom{{\hat{\mu}}_2^{+}}{{\hat{\mu}}_2^{-}}({\mu}_1^{+},{\mu}_1^{-})=
\left(
\begin{array}{cc}
{\hat{\mu}}_2^{+}{\mu}_1^{+} & {\hat{\mu}}_2^{+}{\mu}_1^{-} \\
{\hat{\mu}}_2^{-}{\mu}_1^{+} & {\hat{\mu}}_2^{-}{\mu}_1^{-} \\
 \end{array}
 \right).
\end{equation}
From Eq.(2.18), we know that ${s_{11}}(z)$ is analytical in the upper half plane ${\mathbb{C}_+}$, ${s_{22}}(z)$ is analytical in the lower half plane ${\mathbb{C}_-}$, ${s_{12}}(z)$ and ${s_{21}}(z)$ are continuous to the real axis, but not analytical in the upper and lower half plane.\\
\textbf{Theorem 2.1} The eigenfunctions ${\mu_1}$(z;x,t), ${\mu_2}(z;x,t)$ and spectral matrix function $S(z)$ have the following symmetry property
\begin{equation}
\begin{aligned}
&{\mu_j}(z;x,t)={\sigma_1}\overline{{\mu_j}(\overline{z};x,t)}{\sigma_1},~(j=1,2)\\
&S(z)={\sigma_1}\overline{S(\overline{z})}{\sigma_1},\\
\end{aligned}
\end{equation}
where ${\sigma_1}=\left(
\begin{array}{cc}
0 & 1 \\
1 & 0 \\
 \end{array}
 \right).$\\
\textbf{Proof.}
~~From Lax pair (2.2), we obtain
\begin{equation}
{\mu_{j,x}}(z;x,t)+iz[{\sigma_3,{\mu_j}(z;x,t)}]=P{\mu_j}(z;x,t).
\end{equation}
By replacing $z$ with $\overline{z}$, and taking the conjugate of Eq.(2.20), we get
\begin{equation}
\overline{{\mu_{j,x}}(\overline{z};x,t)}-iz[{\sigma_3},\overline{{\mu_j}(\overline{z};x,t)}]=\overline{P}\overline{{\mu_j}(\overline{z};x,t)}.
\end{equation}
Multiplying the left and right sides of Eq.(2.21) by ${\sigma_1}$ leads to
\begin{equation}
{[{\sigma_1}\overline{{\mu_j}(\overline{z};x,t)}{\sigma_1}]}_x-iz{\sigma_1}{\sigma_3}\overline{{\mu_j}(\overline{z};x,t)}{\sigma_1}
+iz{\sigma_1}\overline{{\mu_j}(\overline{z};x,t)}{\sigma_3}{\sigma_1}={\sigma_1}\overline{P}\overline{{\mu_j}(\overline{z};x,t)}{\sigma_1},
\end{equation}
and
\begin{equation}
{\sigma_1}{\sigma_3}{\sigma_1}=-{\sigma_3},~~~~{\sigma_1}\overline{P}{\sigma_1}=P.
\end{equation}
Therefore, we get
\begin{equation}
{[{\sigma_1}\overline{{\mu_j}(\overline{z};x,t)}{\sigma_1}]}_x+iz[{\sigma_3},{\sigma_1}\overline{{\mu_j}(\overline{z};x,t)}{\sigma_1}]
=P{\sigma_1}\overline{{\mu_j}(\overline{z};x,t)}{\sigma_1}.
\end{equation}
By comparing Eq.(2.2) with Eq.(2.24), we note that ${\mu_j}(z;x,t)$ and ${\sigma_1}\overline{{\mu}_j(\overline{z};x,t)}{\sigma_1}$ satisfy the same differential equation and have the same asymptotic property: ${\mu_j}(z;x,t)$,~~${\sigma_1}\overline{{\mu_j}(\overline{z};x,t)}{\sigma_1}$ $\to I,~x \to \infty$. Consequently,
\begin{equation}
{\mu_j}(z;x,t)={\sigma_1}\overline{{\mu}_j(\overline{z};x,t)}{\sigma_1},
\end{equation}
that is
\begin{equation}
\left(
\begin{array}{cc}
{\mu_{11}}(z) & {\mu_{12}}(z) \\
{\mu_{21}}(z) & {\mu_{22}}(z) \\
 \end{array}
 \right)
 =\left(
\begin{array}{cc}
0 & 1 \\
1 & 0 \\
 \end{array}
 \right)
 \left(
 \begin{array}{cc}
\overline{{\mu_{11}}(\overline{z})} & \overline{{\mu_{12}}(\overline{z})} \\
\overline{{\mu_{21}}(\overline{z})} & \overline{{\mu_{22}}(\overline{z})} \\
 \end{array}
 \right)
 \left(
\begin{array}{cc}
0 & 1 \\
1 & 0 \\
 \end{array}
 \right)
 =\left(
 \begin{array}{cc}
\overline{{\mu_{22}}(\overline{z})} & \overline{{\mu_{21}}(\overline{z})} \\
\overline{{\mu_{12}}(\overline{z})} & \overline{{\mu_{11}}(\overline{z})} \\
 \end{array}
 \right).
\end{equation}
By comparing the two sides of the above equation, we obtain
 ${\mu_{11}}(z)=\overline{{\mu_{22}}(\overline{z})}$,  ${\mu_{12}}(z)=\overline{{\mu_{21}}(\overline{z})}$.

 Nextly, we analyze the symmetry property of $S(z)$. By deforming Eq.(2.7), we have
\begin{equation}
S(z)=e^{(izx+i\alpha z^{3}t)\hat{\sigma}_{3}}\left({\mu_2^{-1}(z;x,t)}{\mu_1}(z;x,t)\right).
\end{equation}
We recall that ${\mu_j}(z;x,t)={\sigma_1}\overline{{\mu}_j(\overline{z};x,t)}{\sigma_1}$, then
\begin{equation}
\begin{aligned}
&{\sigma_1}\overline{S(\overline{z})}{\sigma_1}={\sigma_1}e^{(-izx-i\alpha z^{3}t)\hat{\sigma}_{3}}\overline{{\mu{_2^{-1}}(\overline{z})}}\overline{{\mu_1}(\overline{z})}{\sigma_1}\\
&~~~~~~~~~~~~={\sigma_1}e^{(-izx-i\alpha z^{3}t){\sigma_3}}{\sigma_1}{\sigma_1}\overline{{\mu{_2^{-1}}(\overline{z})}}
{\sigma_1}{\sigma_1}\overline{{\mu_1}(\overline{z})}{\sigma_1}{\sigma_1}e^{(izx+i\alpha z^{3}t){\sigma_3}}{\sigma_1}\\
&~~~~~~~~~~~~=e^{(izx+i\alpha z^{3}t){\sigma_3}}{\mu_2^{-1}}(z){\mu_1}(z)e^{(-izx-i\alpha z^{3}t){\sigma_3}}\\
&~~~~~~~~~~~~=S(z).\\
\end{aligned}
\end{equation}
Therefore, we obtain $\overline{{s_{11}}(\overline{z})}={s_{22}}(z)$ and $\overline{{s_{12}}(\overline{z})}={s_{21}}(z)$.\\
\textbf{Theorem 2.2} The eigenfunction ${\mu_1}(z;x,t)$, ${\mu_2}(z;x,t)$ and spectral matrix function $S(z)$ also have the following symmetry property
\begin{equation}
\begin{aligned}
&{\sigma_3}{\mu}{_j^{H}}(\overline{z};x,t){\sigma_3}={{\mu_j}^{-1}}(z;x,t),\\
&{\sigma_3}{S^{H}}(\overline{z}){\sigma_3}={S^{-1}}(z),\\
\end{aligned}
\end{equation}
where superscript H denotes conjugate transpose and ${\sigma_3}=\left(
\begin{array}{cc}
1 & 0 \\
0 & -1 \\
 \end{array}
 \right)$.\\
\textbf{Proof.}
By direct calculation,
\begin{equation}
\begin{aligned}
&~~~{\sigma_3}{\mu}{_j^{H}}(\overline{z};x,t){\sigma_3}\\
&=\left(
\begin{array}{cc}
1 & 0 \\
0 & -1 \\
 \end{array}
 \right)
{\left(
\begin{array}{cc}
{\mu_{j,11}}(\overline{z},x,t) & {\mu_{j,12}}(\overline{z},x,t) \\
{\mu_{j,21}}(\overline{z},x,t) & {\mu_{j,22}}(\overline{z},x,t) \\
 \end{array}
 \right)}^{H}
 \left(
\begin{array}{cc}
1 & 0 \\
0 & -1 \\
 \end{array}
 \right)\\
 &=\left(
\begin{array}{cc}
1 & 0 \\
0 & -1 \\
 \end{array}
 \right)
 \left(
\begin{array}{cc}
\overline{\mu_{j,11}}(\overline{z},x,t) & \overline{\mu_{j,21}}(\overline{z},x,t) \\
\overline{\mu_{j,12}}(\overline{z},x,t) & \overline{\mu_{j,22}}(\overline{z},x,t) \\
 \end{array}
 \right)
 \left(
\begin{array}{cc}
1 & 0 \\
0 & -1 \\
 \end{array}
 \right)\\
 &=\left(
\begin{array}{cc}
1 & 0 \\
0 & -1 \\
 \end{array}
 \right)
 \left(
\begin{array}{cc}
{\mu_{j,22}}(z) & {\mu_{j,12}}(z) \\
{\mu_{j,21}}(z) & {\mu_{j,11}}(z) \\
 \end{array}
 \right)
 \left(
\begin{array}{cc}
1 & 0 \\
0 & -1 \\
 \end{array}
 \right)\\
 &=\left(
\begin{array}{cc}
{\mu_{j,22}}(z) & -{\mu_{j,12}}(z) \\
-{\mu_{j,21}}(z) & {\mu_{j,11}}(z) \\
 \end{array}
 \right)\\
 &={\mu}{_j^{-1}}(z;x,t).
\end{aligned}
\end{equation}
Therefore,
$\overline{{\mu_{11}}(\overline{z})}={\mu_{22}}(z)$,  $\overline{{\mu_{12}}(\overline{z})}={\mu_{21}}(z)$.
In addition,
\begin{equation}
\begin{aligned}
&~~~{\sigma_3}S^{H}(\overline{z}){\sigma_3}\\
&={\sigma_3}{(e^{(i\overline{z}x+i\alpha \overline{z}^{3}t){\sigma_3}}{\mu}{_2^{-1}}(\overline{z}){\mu_1}(\overline{z})e^{(-i\overline{z}x-i\alpha \overline{z}^{3}t){\sigma_3}})}^{H}{\sigma_3}\\
&={\sigma_3}{(e^{(-izx-i\alpha z^{3}t){\sigma_3}}\overline{{\mu}{_2^{-1}}(\overline{z})}\overline{{\mu_1}(\overline{z})}e^{(izx+i\alpha z^{3}t){\sigma_3}})}^{T}{\sigma_3}\\
&={\sigma_3}e^{(izx+i\alpha z^{3}t){\sigma_3}}{\overline{{\mu_1}(\overline{z})}}^{T}{\overline{{\mu}{_2^{-1}}(\overline{z})}}^{T}e^{(-izx-i\alpha z^{3}t){\sigma_3}}{\sigma_3}\\
&={\sigma_3}e^{(izx+i\alpha z^{3}t){\sigma_3}}{\sigma_3}{\sigma_3}{\mu}{_1^{H}}(\overline{z}){\sigma_3}{\sigma_3}{({\mu}{_2^{-1}}(\overline{z}))}^{H}
{\sigma_3}{\sigma_3}e^{(-izx-i\alpha z^{3}t){\sigma_3}}\\
&=e^{(izx+i\alpha z^{3}t){\sigma_3}}{\mu}{_1^{-1}}(z){\mu_2}(z)e^{(-izx-i\alpha z^{3}t){\sigma_3}}\\
&=S^{-1}(z).\\
\end{aligned}
\end{equation}
So we get $\overline{{s_{11}}(\overline{z})}={s_{22}}(z)$ and $\overline{{s_{12}}(\overline{z})}={s_{21}}(z)$.
Otherwise,
\begin{equation}
S(z)=e^{i(zx+\alpha z^{3}t)\hat{\sigma}_{3}}({\mu_2^{-1}}(z;x,t){\mu_1}(z;x,t)),
\end{equation}
 where ${\mu_2^{-1}}(z;x,t) \to I$ as $t=0$, $x \to +\infty$. Therefore,
\begin{equation}
\begin{aligned}
&S(z)=\lim \limits_{x\to \infty}e^{izx\hat{\sigma}_{3}}{\mu_1}(z;x,0)\\
&~~~~~~=\lim \limits_{x\to \infty}e^{izx\hat{\sigma}_{3}}(I+\int_{-\infty}^{x}e^{-iz(x-\xi)\hat{\sigma}_{3}}P(z;\xi,0){\mu_1}(z;\xi,0)d\xi)\\
&~~~~~~=I+\int_{-\infty}^{\infty}e^{iz\xi\hat{\sigma}_{3}}P(z;\xi,0){\mu_1}(z;\xi,0)d\xi.\\
\end{aligned}
\end{equation}
By calculation,
\begin{equation}
\begin{aligned}
&~~~~S(z)=\left(
\begin{array}{cc}
{s_{11}(z)} & {s_{12}(z)} \\
{s_{21}(z)} & {s_{22}(z)} \\
 \end{array}
 \right)\\
 &=\left(
\begin{array}{cc}
1 & 0 \\
0 & 1 \\
 \end{array}
 \right)
 +\int_{-\infty}^{+\infty}\left(
\begin{array}{cc}
e^{iz\xi} & 0 \\
0 & e^{-iz\xi} \\
 \end{array}
 \right)
 \left(
\begin{array}{cc}
0 & u \\
\overline{u} & 0 \\
 \end{array}
 \right)
 \left(
\begin{array}{cc}
{\mu_{1,11}}(z,x,t) & {\mu_{1,12}}(z,x,t) \\
{\mu_{1,21}}(z,x,t) & {\mu_{1,22}}(z,x,t) \\
 \end{array}
 \right)
 \left(
\begin{array}{cc}
e^{-iz\xi} & 0 \\
0 & e^{iz\xi} \\
 \end{array}
 \right)
 d\xi\\
 &=\left(
\begin{array}{cc}
1 & 0 \\
0 & 1 \\
 \end{array}
 \right)
 +\int_{-\infty}^{+\infty}\left(
\begin{array}{cc}
u{\mu_{1,21}} & ue^{2iz\xi}{\mu_{1,22}} \\
\overline{u}e^{-2iz\xi}{\mu_{1,11}} & \overline{u}{\mu_{1,12}} \\
 \end{array}
 \right)d\xi.\\
\end{aligned}
\end{equation}
Comparing the left and right sides of the above equation, we obtain ${s_{11}(z)}=1+\int_{-\infty}^{+\infty}u{\mu_{1,21}}d\xi$, ${s_{21}(z)}=\int_{-\infty}^{+\infty}\overline{u}e^{-2iz\xi}{\mu_{1,11}}d\xi$.

Then we introduce a piecewise-analytic function $m(z;x,t)$ that
\begin{equation}
m(z;x,t)=
\begin{cases}
    \left(\frac{{\mu}{_1^{+}}}{{s_{11}}},{\mu}{_2^{+}}\right), ~~Imz>0, \\
    \left({\mu}{_2^{-}},\frac{{\mu}{_1^{-}}}{{s_{22}}}\right), ~~Imz<0.
\end{cases}
\end{equation}
\textbf{Theorem 2.3}  The piecewise-analytic function $m(z;x,t)$ defined by Eq.(2.35) satisfies the Riemann-Hilbert problem
\begin{equation}
\begin{aligned}
&\bullet ~~~~{m_\pm}(z;x,t) ~~is ~~analytical ~~in ~~{\mathbb{C}_\pm},\\
&\bullet~~~~{m_+}(z;x,t)={m_-}(z;x,t)v(z;x,t),\\
&\bullet ~~~~{m_\pm}(z;x,t)\to I, as~~ z \to\infty.\\
\end{aligned}
\end{equation}
${m_+}(z;x,t)$ and ${m_-}(z;x,t)$ denote the limiting values as $z$ approaches the contour $\mathbb{R}$ from the left and the right along the contour respectively. The oriented contour on $\mathbb{R}$
 is depicted in Fig. 1, where
\begin{equation}
v(z;x,t)=\left(
\begin{array}{cc}
1-|r(z)|^{2} & -e^{-2it\theta(z)}\overline{r(\overline{z})} \\
e^{2it\theta(z)}{r(z)} & 1 \\
 \end{array}
 \right).
\end{equation}
\textbf{Proof.} From Eq.(2.7), we have
\begin{equation}
\begin{aligned}
&\left[{\mu}{_1^{+}},{\mu}{_1^{-}}\right]=\left[{\mu}{_2^{-}},{\mu}{_2^{+}}\right]\left(
\begin{array}{cc}
e^{-it\theta(z)} & 0 \\
0 & e^{it\theta(z)} \\
 \end{array}
 \right)
\left(
\begin{array}{cc}
{s_{11}}(z) & {s_{12}}(z) \\
{s_{21}}(z) & {s_{22}}(z) \\
 \end{array}
 \right)
 \left(
\begin{array}{cc}
e^{it\theta(z)} & 0 \\
0 & e^{-it\theta(z)} \\
 \end{array}
 \right)\\
 &~~~~~~~~~~~=\left[{s_{11}}(z){\mu}{_2^{-}}+{s_{21}}e^{2it\theta(z)}{\mu}{_2^{+}},{s_{12}}e^{-2it\theta(z)}{\mu}{_2^{-}}+{s_{22}}(z){\mu}{_2^{+}}\right].
\end{aligned}
\end{equation}
By comparing the left and right sides of Eq.(2.38), we get
\begin{equation}
\begin{aligned}
&{\mu}{_1^{+}}={s_{11}}(z){\mu}{_2^{-}}+{s_{21}}(z)e^{2it\theta(z)}{\mu}{_2^{+}},\\ &{\mu}{_1^{-}}={s_{12}}(z)e^{-2it\theta(z)}{\mu}{_2^{-}}+{s_{22}}(z){\mu}{_2^{+}}.\\
\end{aligned}
\end{equation}
Deforming the above equation, we have
\begin{equation}
\begin{aligned}
&\frac{{\mu}{_1^{+}}}{{s_{11}}(z)}=\frac{1}{{s_{11}(z)}{s_{22}}(z)}{\mu}{_2^{-}}+\frac{1}{{s_{22}}(z)}\frac{{s_{21}}(z)}{{s_{11}}(z)}e^{2it\theta(z)}{\mu}{_1^{-}},\\
&{\mu}{_2^{+}}=\frac{{\mu}{_1^{-}}}{{s_{22}}(z)}-\frac{{s_{12}}(z)}{{s_{22}}(z)}e^{-2it\theta(z)}{\mu}{_2^{-}}.\\
\end{aligned}
\end{equation}
The above system in matrix form is
\begin{equation}
\left[\frac{{\mu}{_1^{+}}}{{s_{11}}(z)},{\mu}{_2^{+}}\right]=\left[{\mu}{_2^{-}},\frac{{\mu}{_1^{-}}}{{s_{22}}(z)}\right]e^{-it\theta(z)\hat{\sigma}_{3}}
\left(
\begin{array}{cc}
\frac{1}{{s_{11}}(z){s_{22}}(z)} & -\frac{{s_{12}}(z)}{{s_{22}}(z)} \\
\frac{{s_{21}}(z)}{{s_{11}}(z)} & 1 \\
 \end{array}
 \right).
\end{equation}
Let $r(z)=\frac{{s_{21}(z)}}{{s_{11}(z)}}$, when $z\in \mathbb{R}$ we have
\begin{equation}
\overline{{s_{11}}(\overline{z})}={s_{22}}(z),\overline{{s_{12}}(\overline{z})}={s_{21}}(z),
\end{equation}
\begin{equation}
-\frac{{s_{12}}(z)}{{s_{22}}(z)}=-\frac{\overline{s_{21}(\overline{z})}}{\overline{s_{11}(\overline{z})}}
=-\overline{r(\overline{z})},
\end{equation}
\begin{equation}
\frac{1}{{s_{11}}(z){s_{22}}(z)}=\frac{{s_{11}}(z){s_{22}}(z)-{s_{21}}(z){s_{12}}(z)}{{s_{11}}(z){s_{22}}(z)}=1-\frac{{s_{21}}(z)}{{s_{11}}(z)}
\frac{{s_{12}}(z)}{{s_{22}}(z)}=1-r(z)\overline{r(\overline{z})}.
\end{equation}
From Eq.(2.15), we note that
\begin{equation}
\det(S(z))={s_{11}}(z){s_{22}}(z)-{s_{21}}(z){s_{12}}(z)=1.
\end{equation}
By dividing both sides of Eq.(2.45) by ${s_{11}}(z){s_{22}}(z)$ and using Eq.(2.42), we obtain
\begin{equation}
r(z)\overline{r(\overline{z})}+\frac{1}{{s_{11}}(z)\overline{{s_{11}}(\overline{z})}}=1.
\end{equation}
When $z \in R$, we have
\begin{equation}
|r(z)|^2+\frac{1}{|{s_{11}}(z)|^2}=1,
\end{equation}
which means $|r(z)|<1$. By using Eq.(2.42)-(2.44), Eq.(2.41) can be turned into
\begin{equation}
\left[\frac{{\mu}{_1^{+}}}{{s_{11}}(z)},{\mu}{_2^{+}}\right]=\left[{\mu}{_2^{-}},\frac{{\mu}{_1^{-}}}{{s_{22}}(z)}\right]e^{-it\theta(z)\hat{\sigma}_{3}}
\left(
\begin{array}{cc}
1-r(z)\overline{r(\overline{z})} & -\overline{r(\overline{z})} \\
r(z) & 1 \\
 \end{array}
 \right).
\end{equation}
The solution of Riemann-Hilbert problem (2.36) exists and is unique, the function
\begin{equation}
u(x,t)=\lim \limits_{x\to \infty}{{(z\mu(z;x,t))}_{12}},
\end{equation}
solves complex mKdV equation.\\

\begin{figure}
  \centering
  \includegraphics[width=8cm]{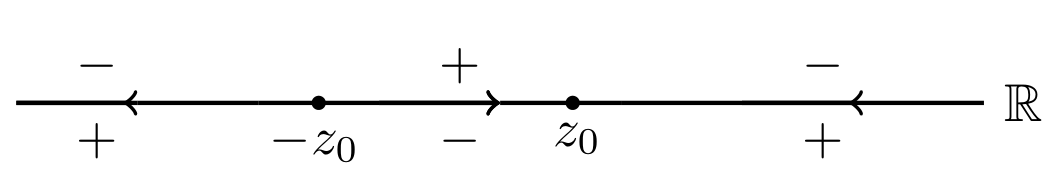}
  \caption{The oriented contour ${\Sigma^{(1)}}$.}
\end{figure}

\section{Long-Time Asymptotic Behavior}
\subsection{Factorization of the jump matrix}
Notice that there are two oscillatory terms $e^{it\theta(z)}$ and $e^{-it\theta(z)}$ in the jump matrix $v(z;x,t)$, we take
\begin{equation}
\varphi(z)=i\theta(z)=i\left(\frac{x}{t}z+\alpha z^{3}\right),
\end{equation}
then
\begin{equation}
\varphi^{'}(z)=i\frac{x}{t}+3i\alpha z^{2},  \varphi^{''}(z)=6i\alpha z\neq0.
\end{equation}
By assuming that $\varphi^{'}(z)=i\frac{x}{t}+3i\alpha z^{2}=0$, we get two steady state phase point $\pm{z_0}=\pm\sqrt{-\frac{x}{3\alpha t}}$. Substituting the expression of ${z_0}$ into $\theta(z)$, we have
\begin{equation}
\theta(z)=z\frac{x}{t}+\alpha z^{3}=z\left(-3\alpha {z_0^{2}}\right)+\alpha z^{3}=\alpha\left(z^{3}-3{z_0^{2}}z\right).
\end{equation}
Via deforming Eq.(3.3), we get
\begin{equation}
\theta(z)=\alpha\left((z+{z_0})^{3}-3{z_0}(z+{z_0})^{2}+2{z_0^3}\right),
\end{equation}
and
\begin{equation}
\theta(z)=\alpha\left((z-{z_0})^{3}+3{z_0}(z-{z_0})^{2}-2{z_0^3}\right).
\end{equation}
The jump matrix $v(z;x,t)$ has the lower/upper triangular factorization
\begin{equation}
v(z;x,t)=\left(
\begin{array}{cc}
1 & -\overline{r(\overline{z})}e^{-2it\theta(z)} \\
0 & 1 \\
 \end{array}
 \right)
 \left(
\begin{array}{cc}
1 & 0 \\
r(z)e^{2it\theta(z)} & 1 \\
 \end{array}
 \right),~~~(z\to \infty)
\end{equation}
and the upper/diagonal/lower factorization
\begin{equation}
v(z;x,t)=\left(
\begin{array}{cc}
1 & 0 \\
e^{2it\theta(z)}\frac{r(z)}{1-r(z)\overline{r(\overline{z})}} & 1 \\
 \end{array}
 \right)
 \left(
\begin{array}{cc}
1-r(z)\overline{r(\overline{z})} & 0 \\
0 & \frac{1}{1-r(z)\overline{r(\overline{z})}} \\
 \end{array}
 \right)
 \left(
\begin{array}{cc}
1 & -e^{-2it\theta(z)}\frac{\overline{r(\overline{z})}}{1-r(z)\overline{r(\overline{z})}} \\
0 & 1 \\
 \end{array}
 \right),~~~z\in(-{z_0},{z_0}).
\end{equation}
Now we seek a transformation to eliminate the middle diagonal matrix in $z\in\left(-{z_0},{z_0}\right)$ factorization. Making a transformation
\begin{equation}
m^{(1)}=m\delta^{-{\sigma_3}},
\end{equation}
then
\begin{equation}
{m}{_+^{(1)}}={m_+}{\delta}{_+^{-{\sigma_3}}}={m_-}v(z;x,t){\delta}{_+^{-{\sigma_3}}}={m_-}{\delta}{_-^{-{\sigma_3}}}
{\delta}{_+^{{\sigma_3}}}v(z;x,t){\delta}{_+^{-{\sigma_3}}}={m}{_-^{(1)}}{\delta}{_-^{{\sigma_3}}}v(z;x,t)
{\delta}{_+^{-{\sigma_3}}}={m}{_-^{(1)}}{v^{(1)}}(z;x,t),
\end{equation}
therefore,
\begin{equation}
{v^{(1)}}(z;x,t)={\delta}{_-^{{\sigma_3}}}v(z;x,t){\delta}{_+^{-{\sigma_3}}}.
\end{equation}
For $z\in\left(-{z_0},{z_0}\right)$,
\begin{equation}
\begin{aligned}
&~~~~{v^{(1)}}(z;x,t)\\
&={\delta}{_-^{{\sigma_3}}}\left(
\begin{array}{cc}
1 & 0 \\
e^{2it\theta(z)}\frac{r(z)}{1-r(z)\overline{r(\overline{z})}} & 1 \\
 \end{array}
 \right)
 \left(
\begin{array}{cc}
1-r(z)\overline{r(\overline{z})} & 0 \\
0 & \frac{1}{1-r(z)\overline{r(\overline{z})}} \\
 \end{array}
 \right)
 \left(
\begin{array}{cc}
1 & -e^{-2it\theta(z)}\frac{\overline{r(\overline{z})}}{1-r(z)\overline{r(\overline{z})}} \\
0 & 1 \\
 \end{array}
 \right)
 {\delta}{_+^{-{\sigma_3}}}\\
 &=\left(
\begin{array}{cc}
1 & 0 \\
{\delta}{_-^{-2}}e^{2it\theta(z)}\frac{r(z)}{1-r(z)\overline{r(\overline{z})}} & 1 \\
 \end{array}
 \right)
\left(
\begin{array}{cc}
\left(1-r(z)\overline{r(\overline{z})}\right){\delta_-}{\delta}{_+^{-1}} & 0 \\
0 & \left(\frac{1}{1-r(z)\overline{r(\overline{z})}}\right){\delta}{_-^{-1}}{\delta_+} \\
 \end{array}
 \right)
 \left(
\begin{array}{cc}
1 & -{\delta}{_+^{2}}e^{-2it\theta(z)}\frac{\overline{r(\overline{z})}}{1-r(z)\overline{r(\overline{z})}} \\
0 & 1 \\
 \end{array}
 \right).\\
\end{aligned}
\end{equation}
For $z\to \infty$,
\begin{equation}
\begin{aligned}
&~~~~{v^{(1)}}(z;x,t)\\
&={\delta}{_-^{{\sigma_3}}}\left(
\begin{array}{cc}
1 & -\overline{r(\overline{z})}e^{-2it\theta(z)} \\
0 & 1 \\
 \end{array}
 \right)
 \left(
\begin{array}{cc}
1 & 0 \\
r(z)e^{2it\theta(z)} & 1 \\
 \end{array}
 \right)
 {\delta}{_+^{-{\sigma_3}}}\\
 &=\left(
\begin{array}{cc}
1 & -{\delta}{_-^{2}}\overline{r(\overline{z})}e^{-2it\theta(z)} \\
0 & 1 \\
 \end{array}
 \right)
\left(
\begin{array}{cc}
1 & 0 \\
{\delta}{_-^{-2}}r(z) e^{2it\theta(z)}  & 1 \\
 \end{array}
 \right).
\end{aligned}
\end{equation}
Now we introduce a scale Riemann-Hilbert problem
\begin{equation}
\begin{aligned}
&\bullet \delta(z) ~~is ~~analytical~~ in ~~\mathbb{C} \backslash \mathbb{R} ,\\
&\bullet {\delta_+}(z)={\delta_-}(z)\left(1-r(z)\overline{r(\overline{z})}\right),~~z\in\left(-{z_0},{z_0}\right),\\
&\bullet {\delta_+}(z)={\delta_-}(z),~~z\to \infty,\\
&\bullet \delta(z)\to 1,~~z\to \infty.\\
\end{aligned}
\end{equation}
By deforming the second condition of the above Riemann-Hilbert problem, we get
\begin{equation}
{\left(\log\delta(z)\right)_+}-{\left(\log\delta(z)\right)_-}=f(z)=\begin{cases}
\log\left(1-r(z)\overline{r(\overline{z})}\right), z\in\left(-{z_0},{z_0}\right)\\
0, ~~~~~~~~~~~~~~~~~~ z\to \infty\\
\end{cases}.
\end{equation}
According to Plemelj formula, Eq.(3.14) has a unique bounded solution
\begin{equation}
\log\delta(z)=\frac{1}{2\pi i}\int_{-\infty}^{+\infty}\frac{f(\xi)}{\xi-z}d\xi=\frac{1}{2\pi i}\int_{-\infty}^{{z_0}}\frac{\log(1-r(z)r(\overline{z}))}{\xi-z}d\xi.
\end{equation}
Then we calculate the right end of Eq.(3.15),
\begin{equation}
\begin{aligned}
&~~~~\frac{1}{2\pi i}\int_{-{z_0}}^{{z_0}}\frac{\ln(1-|r(z)|^{2})}{\xi-z}d\xi\\
&=\frac{1}{2\pi i}\int_{-{z_0}}^{{z_0}}\ln\left(\frac{1-|r(\xi)|^{2}}{1-|r({z_0})|^{2}}\right)\frac{d\xi}{\xi-z}+\frac{1}{2\pi i}\int_{-{z_0}}^{{z_0}}\ln\left(1-|r({z_0})|^{2}\right)\frac{d\xi}{\xi-z}\\
&=\frac{1}{2\pi i}\int_{-{z_0}}^{{z_0}}\ln\left(\frac{1-|r(\xi)|^{2}}{1-|r({z_0})|^{2}}\right)\frac{d\xi}{\xi-z}+
\frac{1}{2\pi i}\ln\left(1-|r({z_0})|^{2}\right)\int_{-{z_0}}^{{z_0}}\frac{d\xi}{\xi-z}\\
&=\frac{1}{2\pi i}\int_{-{z_0}}^{{z_0}}\ln\left(\frac{1-|r(\xi)|^{2}}{1-|r({z_0})|^{2}}\right)\frac{d\xi}{\xi-z}+
\frac{1}{2\pi i}\ln\left(1-|r({z_0})|^{2}\right)\ln\left(\frac{z-{z_0}}{z+{z_0}}\right),\\
\end{aligned}
\end{equation}
therefore,
\begin{equation}
\delta(z)=\left(\frac{z-{z_0}}{z+{z_0}}\right)^{\frac{1}{2\pi i}\ln\left(1-|r({z_0})|^{2}\right)}e^{\frac{1}{2\pi i}\int_{-{z_0}}^{{z_0}}\ln\left(\frac{1-|r(\xi)|^{2}}{1-|r({z_0})|^{2}}\right)\frac{d\xi}{\xi-z}}.
\end{equation}
For convenience, we write $\delta(z)$ in the following form
\begin{equation}
\delta(z)=\left(\frac{z-{z_0}}{z+{z_0}}\right)^{i\upsilon}e^{\chi(z)},
\end{equation}
where $\upsilon=-\frac{1}{2\pi}\ln\left(1-|r({z_0})|^{2}\right)$ and
$\chi(z)=\frac{1}{2\pi i}\int_{-{z_0}}^{{z_0}}\ln\left(\frac{1-|r(\xi)|^{2}}{1-|r({z_0})|^{2}}\right)\frac{d\xi}{\xi-z}$.\\

Based on uniqueness, we derive that
\begin{equation}
\delta(z)=\left(\overline{\delta(\overline{z})}\right)^{-1}.
\end{equation}
Substituting Eq.(3.19) into Eq.(3.13), for $z \in R$ we obtain
\begin{equation}
|\delta_{-}(z)|^{2}=
\begin{cases}
    \frac{1}{1-|r|^{2}}, ~~|z|<z_{0}, \\
    2, ~~|z|>z_{0},
\end{cases}
\end{equation}
and
\begin{equation}
|\delta_{+}(z)|^{2}=
\begin{cases}
    1-|r|^{2}, ~~|z|<z_{0}, \\
    2, ~~|z|>z_{0}.
\end{cases}
\end{equation}
According to the maximum principle,
\begin{equation}
|\delta(z)|\leq c <\infty,
\end{equation}
for $z \in \mathbb{C}$. Similarly, $\delta(z)$ has a similar process in $z \in (-\infty,{z_0})$. On the basis of the above analysis, $m^{(1)}$ satisfy a new Riemann-Hilbert problem
\begin{equation}
\begin{aligned}
&\bullet m^{(1)}(z;x,t)~~is ~~analytical  ~~in ~~\mathbb{C} \backslash \Sigma^{(1)},\\
&\bullet {m_+^{(1)}}(z;x,t)={m_-^{(1)}}(z;x,t)v^{(1)}(z;x,t),~~~z\in \Sigma^{(1)},\\
&\bullet m^{(1)}(z;x,t) \to I, ~~as ~z\to \infty.\\
\end{aligned}
\end{equation}
The jump matrix $v^{(1)}(z;x,t)$ is
\begin{equation}
v^{(1)}(z;x,t)=\begin{cases}
\left(
\begin{array}{cc}
1 & 0 \\
{\delta}{_-^{-2}}e^{2it\theta(z)}\frac{r(z)}{1-r(z)\overline{r(\overline{z})}} & 1 \\
 \end{array}
 \right)
\left(
\begin{array}{cc}
1 & -{\delta}{_+^{2}}e^{-2it\theta(z)}\frac{\overline{r(\overline{z})}}{1-r(z)\overline{r(\overline{z})}} \\
0 & 1 \\
 \end{array}
 \right), ~~z\in(-{z_0},{z_0}),\\
 \left(
\begin{array}{cc}
1 & -{\delta}{_-^{2}}\overline{r(\overline{z})}e^{-2it\theta(z)} \\
0 & 1 \\
 \end{array}
 \right)
\left(
\begin{array}{cc}
1 & 0 \\
{\delta}{_-^{-2}}r(z) e^{2it\theta(z)}  & 1 \\
 \end{array}
 \right),~~z\to \infty,
\end{cases}
\end{equation}
and we have
\begin{equation}
u(x,t)=2i\lim \limits_{z\to \infty}{\left(zm^{(1)}\left(z;x,t\right)\right)_{12}}.
\end{equation}
For the sake of convenience, we introduce a function
\begin{equation}
\rho(z)=
\begin{cases}
\overline{r(\overline{z})}, ~~~~~~~~z \to \infty,\\
\frac{-\overline{r(\overline{z})}}{1-r(z)\overline{r(\overline{z})}},~~~z\in (-{z_0},{z_0}),
\end{cases}
\end{equation}
then the jump matrix $v^{(1)}(z;x,t)$ is unified as
\begin{equation}
v^{(1)}(z;x,t)={\left(
\begin{array}{cc}
1 & 0 \\
{\delta}{_-^{-2}}e^{2it\theta(z)}\overline{\rho(\overline{z})} & 1 \\
 \end{array}
 \right)}^{-1}
 \left(
\begin{array}{cc}
1 & {\delta}{_+^{2}}e^{-2it\theta(z)}\rho(z) \\
0 & 1 \\
 \end{array}
 \right).
\end{equation}

\subsection{ Extend to the augmented contour}
Let $L$ denote the contour
\begin{equation}
\begin{aligned}
&L:{z={z_0}+{z_0}\beta e^{\frac{3\pi i}{4}}:-\infty<\beta\leq\sqrt{2}}\\
&\cup {z=-{z_0}+{z_0}\beta e^{\frac{\pi i}{4}}:-\infty<\beta\leq\sqrt{2}},\\
\end{aligned}
\end{equation}
and
\begin{equation}
\begin{aligned}
&{L_\varepsilon}:{z={z_0}+{z_0}\beta e^{\frac{3\pi i}{4}}:\varepsilon<\beta\leq\sqrt{2}}\\
&\cup {z=-{z_0}+{z_0}\beta e^{\frac{\pi i}{4}}:\varepsilon<\beta\leq\sqrt{2}},
\end{aligned}
\end{equation}
where $0<\varepsilon<\sqrt{2}$.\\
Through the analysis of $Re(i\theta(z))$, we get the symbol distribution map of $Re(i\theta(z))$, the map is shown in Fig.2.

\begin{figure}
  \centering
  \includegraphics[width=8cm]{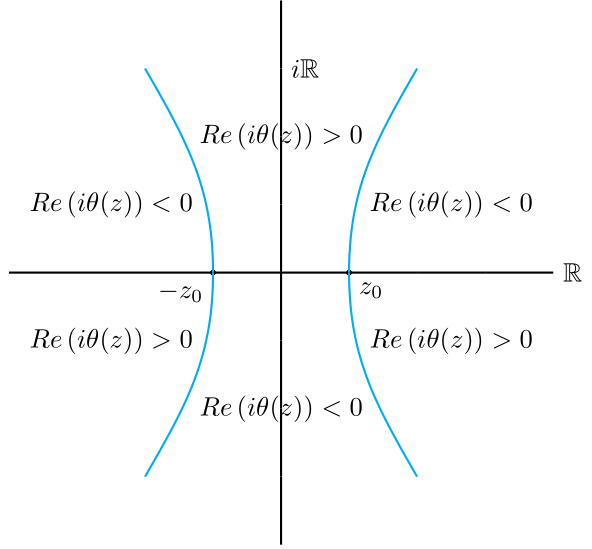}
  \caption{Symbol distribution map of $Re(i\theta(z))$.}
\end{figure}

\textbf{Theorem 3.1} The spectral function $\rho(z)$ has a decomposition on the real number axis as follows
\begin{equation}
\rho(z)=R(z)+{h_1}(z)+{h_2}(z),
\end{equation}
 $R(z)$, ${h_1}(z)$ and ${h_2}(z)$ satisfy
\begin{equation}
|e^{-2it\theta(z)}{h_1}(z)|\lesssim \frac{1}{|z-i|^{2}t^{l}},~~z \in \mathbb{R},
\end{equation}
\begin{equation}
|e^{-2it\theta(z)}{h_2}(z)|\lesssim \frac{1}{|z-i|^{2}t^{l}},~~z \in L,
\end{equation}
\begin{equation}
|e^{-2it\theta(z)}R(z)|\lesssim ce^{-6\alpha \varepsilon^{2}\tau},~~z \in {L_\varepsilon},
\end{equation}
where $\theta(z)=\alpha\left(z-{z_0}\right)^{3}+3\alpha {z_0}\left(z-{z_0}\right)^{2}-2{z_0}^{3}$, $R(z)$ is piecewise rational, ${h_2}(z)$ has an analytic continuation to $L$, $l$ is a arbitrary positive integer.\\
\textbf{Proof.} ~Firstly, we consider spectral function $\rho(z)$ with $z \in \left({z_0},+\infty\right)$. Replace the function $\rho(z)$ by a rational function with well-controlled errors, expand $\left(z-i\right)^{m+5}\rho(z)$ in a Taylor series around ${z_0}$
\begin{equation}
\begin{aligned}
&~~~~\left(z-i\right)^{m+5}\rho(z)\\
&=\left({z_0}-i\right)^{m+5}\rho({z_0})+\frac{\left((z-i)^{m+5}\rho(z)\right)^{(1)}}{1}{\mid_{z={z_0}}}(z-{z_0})+\ldots+
\frac{\left((z-i)^{m+5}\rho(z)\right)^{(m)}}{m!}{\mid_{z={z_0}}}(z-{z_0})^{m}\\
&~~~~+\frac{\left((\xi-i)^{m+5}\rho(\xi)\right)^{(m+1)}}{(m+1)!}(z-{z_0})^{m+1}\\
&={\mu_0}+{\mu_1}(z-{z_0})+\ldots+{\mu_m}(z-{z_0})^{m}+\frac{1}{m!}\int_{{z_0}}^{z}\left((\xi-i)^{m+5}\rho(\xi)\right)^{m+1}
(z-\xi)^{m}d\xi.\\
\end{aligned}
\end{equation}
Take
\begin{equation}
R(z)=\frac{\sum_{j=0}^{m}{\mu_j}\left(z-{z_0}\right)^{j}}{\left(z-i\right)^{m+5}}
\end{equation}
and
\begin{equation}
h(z)=\rho(z)-R(z),
\end{equation}
then
\begin{equation}
h(z)=\frac{\left((\xi-i)^{m+5}\rho(\xi)\right)^{(m+1)}}{(z-i)^{m+5}m!}\left(z-{z_0}\right)^{m+1}.
\end{equation}
We have
\begin{equation}
\frac{d^{j}h(z)}{{dz}^{j}}\mid_{z={z_0}}=0,~~0\leq j\leq m,
\end{equation}
so
\begin{equation}
\frac{d^{j}\rho(z)}{{dz}^{j}}\mid_{z={z_0}}=\frac{d^{j}R(z)}{{dz}^{j}}\mid_{z={z_0}},~~0\leq j\leq m.
\end{equation}
 From $\theta(z)=\frac{z}{t}x+\alpha z^{3}$ and ${z_0}=\sqrt{\frac{-x}{3\alpha t}}$, we have $-x=3\alpha t{z_0}^{2}$, therefore
\begin{equation}
t\theta(z)=zx+\alpha z^{3}t=-x\left(-z+\frac{\alpha tz^{3}}{-x}\right)=|x|\left(-z+\frac{\alpha tz^{3}}{3\alpha t{z_0}^{2}}\right)=|x|\left(\frac{z^{3}-3{z_0}^{2}z}{3{z_0}^{2}}\right)=|x|\tilde{\theta}(z).
\end{equation}
Then $Re(i\theta(z))$ and $Re(i\tilde{\theta}(z))$ differ by a constant multiplier. Next we define
\begin{equation}
\gamma(z)=\frac{(z-{z_0})^{n}}{(z-i)^{n+2}},
\end{equation}
by Fourier inverse transformation,
\begin{equation}
\left(\frac{h}{\gamma}\right)(z)=\int_{-\infty}^{+\infty}e^{is\theta(z)}\hat{\left(\frac{h}{\gamma}\right)}(s)\bar{d}s,
\end{equation}
where
\begin{equation}
\hat{\left(\frac{h}{\gamma}\right)}(s)=\int_{{z_0}}^{+\infty}e^{-is\theta(z)}\left(\frac{h}{\gamma}\right)(z)\bar{d}\theta(z)
\end{equation}
and
\begin{equation}
\bar{d}s=\frac{ds}{\sqrt{2\pi}},~~\bar{d}\theta(z)=\frac{d\theta(z)}{\sqrt{2\pi}}.
\end{equation}
Due to
\begin{equation}
h(z)=\frac{1}{(z-i)^{m+5}m!}\int_{{z_0}}^{z}\left((\xi-i)^{m+5}\rho(\xi)\right)^{(m+1)}(z-\xi)^{m}d\xi,
\end{equation}
then
\begin{equation}
\left(\frac{h}{\gamma}\right)(z)=\frac{(z-i)^{n+2}\left(z-{z_0}\right)^{-n}}{(z-i)^{m+5}m!}\int_{{z_0}}^{z}((\xi-i)^{m+5}\rho(\xi))^{(m+1)}\left(z-\xi\right)^{m}d\xi.
\end{equation}
By taking $\xi={z_0}+\eta(z-{z_0})$,
\begin{equation}
\begin{aligned}
&\left(\frac{h}{\gamma}\right)(z)=\frac{(z-{z_0})^{-n}}{(z-i)^{m+3-n}}\frac{1}{m!}\int_{0}^{1}\left[\left({z_0}+\eta(z-{z_0})-i\right)^{m+5}\rho\left({z_0}+\eta(z-{z_0})\right)\right]^{(m+1)}\\
&~~~~~~~~~~~~~\cdot\left(z-{z_0}-\eta(z-{z_0})\right)^{m}d\left({z_0}+\eta(z-{z_0})\right)\\
&~~~~~~~~~~~~=\frac{\left(z-{z_0}\right)^{m+1-n}}{\left(z-i\right)^{m+3-n}}\frac{1}{m!}\int_{0}^{1}[\left({z_0}+\eta(z-{z_0})-i\right)^{m+5}\rho\left({z_0}+\eta\left(z-{z_0}\right)\right)]^{(m+1)}(1-\eta)^{m}d\eta\\
&~~~~~~~~~~~~=\frac{\left(z-{z_0}\right)^{m+1-n}}{\left(z-i\right)^{m+3-n}}\frac{1}{m!}g(z,{z_0}).
\end{aligned}
\end{equation}
Let $m=4n+1$, then
\begin{equation}
\left(\frac{h}{\gamma}\right)(z)=\frac{\left(z-{z_0}\right)^{3n+2}}{(z-i)^{3n+4}}g\left(z,{z_0}\right),
\end{equation}
where $g\left(z,{z_0}\right)=\frac{1}{m!}\int_{0}^{1}\left[\left({z_0}+\eta(z-{z_0})-i\right)^{m+5}\rho\left({z_0}+\eta(z-{z_0})\right)\right]^{(m+1)}\left(1-\eta\right)^{m}d\eta$.
Due to $\rho(z)\in \mathcal{S}(R)$, we have
\begin{equation}
\left|\frac{d^{j}g(z,{z_0})}{{dz}^{j}}\right|\lesssim c,~~z\geqslant {z_0}.
\end{equation}
Otherwise,
\begin{equation}
\left|\frac{z-{z_0}}{z+{z_0}}\right|\leq 1, ~~z\geqslant {z_0},
\end{equation}
we have
\begin{equation}
\begin{aligned}
&\int_{{z_0}}^{\infty}\left|\left(\frac{d}{d\theta}\right)^{j}\left(\frac{h}{\gamma}\right)(z)\right|^{2}\bar{d}\theta(z)\\
&=\int_{{z_0}}^{\infty}\left|\left(\frac{1}{3\alpha (z^{2}-{z_0}^{2})}\right)^{j}\left(\frac{h}{\gamma}\right)(z)\right|^{2}3\alpha \left(z^{2}-{z_0}^{2}\right)\bar{d}z\\
&\leq c\int_{{z_0}}^{\infty}\left|\frac{\left(z-{z_0}\right)^{3n+2-3j}}{(z-i)^{3n+4}}\right|^{2}\left(z^{2}-{z_0}^{2}\right)\bar{d}z\leq c, ~~\left(0<{z_0}<M,~0\leq j\leq \frac{3n+2}{3}\right).
\end{aligned}
\end{equation}
By using Plancherel theorem, we have
\begin{equation}
\int_{-\infty}^{+\infty}\left(1+s^{2}\right)^{j}\left|\hat{\left(\frac{h}{\gamma}\right)}(s)\right|^{2}\leq c \leq \infty,~~\left(0<{z_0}<M,~0\leq j\leq \frac{3n+2}{3}\right).
\end{equation}
Next we split $h(z)$
\begin{equation}
\begin{aligned}
&h(z)=\gamma(z)\int_{-\infty}^{+\infty}e^{is\theta(z){\hat{\left(\frac{h}{\gamma}\right)}(s)}}\bar{d}s\\
&~~~~~~=\gamma(z)\int_{t}^{+\infty}e^{is\theta(z){\hat{\left(\frac{h}{\gamma}\right)}(s)}}\bar{d}s+\gamma(z)\int_{-\infty}^{t}e^{is\theta(z){\hat{\left(\frac{h}{\gamma}\right)}(s)}}\bar{d}s\\
&~~~~~~={h_1}(z)+{h_2}(z).
\end{aligned}
\end{equation}
According to Eq.(3.41), we have
\begin{equation}
\left|e^{-2it\theta(z)}{h_1}(z)\right|=\left|e^{-2i|x|\tilde{\theta}(z)}{h_1}(z)\right|.
\end{equation}
 When $z\geq {z_0}\in \mathbb{R}$, $Re(i\tilde{\theta}(z))=0$, then $-2i|x|\tilde{\theta}(z)$
is pure imaginary, so
\begin{equation}
\begin{aligned}
&~~~\left|e^{-2i|x|\tilde{\theta}(z)}{h_1}(z)\right|\\
&\leq \left|\gamma(z)\right|\int_{t}^{+\infty}\left|{\hat{\left(\frac{h}{\gamma}\right)}(s)}\right|\bar{d}s\\
&= \left|\gamma(z)\right|\int_{t}^{+\infty}\left(1+s^{2}\right)^{-\frac{p}{2}}\left(1+s^{2}\right)^{\frac{p}{2}}\left|{\hat{\left(\frac{h}{\gamma}\right)}(s)}\right|\bar{d}s\\
&\leq\left|\gamma(z)\right|\left(\int_{t}^{+\infty}(1+s^{2})^{-p}ds\right)^{\frac{1}{2}}\left(\int_{t}^{+\infty}(1+s^{2})^{p}\left|{\hat{\left(\frac{h}{\gamma}\right)}(s)}\right|^{2}\bar{d}s\right)^{\frac{1}{2}}\\
&\lesssim \left|\gamma(z)\right|\left(\int_{t}^{+\infty}s^{-2p}ds\right)^{\frac{1}{2}}\\
&\lesssim \frac{1}{\left|z-i\right|^{2}}t^{l}.
\end{aligned}
\end{equation}
For $\left|e^{-2i|x|\tilde{\theta}(z)}{h_2}(z)\right|$, on ray ${z={z_0}+{z_0}\eta e^{-\frac{i\pi}{4}}}$, we have
\begin{equation}
\begin{aligned}
&~~~\left|e^{-2i|x|\tilde{\theta}(z)}{h_2}(z)\right|\\
&\leq \left|e^{-2it\theta(z)}\right|\cdot\left|\frac{(z-{z_0})^{n}}{(z-i)^{n+2}}\right|\cdot\left|\int_{-\infty}^{t}e^{is\theta(z)}{\hat{\left(\frac{h}{\gamma}\right)}(s)}\right|\bar{d}s\\
&\leq c \left|\frac{(z-{z_0})^{n}}{(z-i)^{n+2}}\right|\cdot\left| e^{-tRe(i\theta(z))}\right|\cdot\left|\int_{-\infty}^{t}e^{i(s-t)\theta(z)}{\hat{\left(\frac{h}{\gamma}\right)}(s)}\right|\bar{d}s\\
&\leq c\left|\frac{(z-{z_0})^{n}}{(z-i)^{n+2}} \right| \cdot e^{-tRe(i\theta(z))}\cdot(\int_{-\infty}^{t}(1+s^{2})^{\frac{1}{2}})
\cdot\int_{-\infty}^{t}((1+s^{2})^{2}\left|\left(\frac{h}{\gamma}\right)(s)\right|^{2}ds)^{\frac{1}{2}}\\
&\leq \frac{{z_0}^{n}u^{n}}{\left|z-i\right|^{n+2}}\cdot e^{-tRe(i\theta(z))}\\
&\leq\frac{c{z_0^{n}}\left[((t{z_0^{3}})^{\frac{1}{3}}u)^{n}\cdot e^{-\frac{\sqrt{2}}{2}}\alpha {z_0^{3}}u^{3}t\right]}{\left|z-i\right|^{n+2}\cdot(t{z_0^{3}})^{\frac{n}{3}}}\\
&\leq \frac{c}{\left|z-i\right|^{2}t^{\frac{n}{3}}} \leq \frac{c}{\left|z-i\right|^{2}t^{l}},\\
\end{aligned}
\end{equation}
\begin{equation}
\left|e^{-2it\theta(z)}\left[\rho(z)\right]\right|\leq ce^{-2tRe(i\theta(z))}\leq ce^{-6\alpha t{z_0^{3}}u^{2}}\leq ce^{-6\alpha\tau},
\end{equation}
where $\tau=t{z_0^{3}}=\left(\frac{-x}{3\alpha t^{\frac{1}{3}}}\right)^{\frac{3}{2}}$. \\
For $\overline{\rho(\overline{z})}$, by taking Schwartz conjugate to $\rho(z)$, we have
\begin{equation}
\overline{\rho(\overline{z})}=\overline{R(\overline{z})}+\overline{{h_1}(\overline{z})}+\overline{{h_2}(\overline{z})}.
\end{equation}
There are similar estimates for $e^{2it\theta(z)}\overline{{h_1}(\overline{z})}$, $e^{2it\theta(z)}\overline{{h_2}(\overline{z})}$ and $e^{2it\theta(z)}\overline{R(\overline{z})}$ on $\mathbb{R}\cup {\overline{L}}$. Based on the result of theorem 3.1 and Beal-Cofiman theorem, $v^{(1)}(z;x,t)$ has the following decomposition
\begin{equation}
 v^{(1)}(z;x,t)=({b_-})^{-1}{b_+}=({b_-^{a}})^{-1}({b_-^{o}})^{-1}{b_+^{o}}{b_+^{a}},
  \end{equation}
 ${b_\pm}$ has decomposition
\begin{equation}
\begin{aligned}
&{b_+}={b}{_+^{o}}{b}{_+^{a}}=(I+{\omega}{_+^{o}})(I+{\omega}{_+^{a}}),\\
&{b_-}={b}{_-^{o}}{b}{_-^{a}}=(I-{\omega}{_-^{o}})(I-{\omega}{_-^{a}}),\\
\end{aligned}
\end{equation}
where\\
$~~~~~~~~~~~~~~~~~~~~{\omega_+^{o}}=\left(
\begin{array}{cc}
0 & {\delta}{_+^{2}}(z)e^{-2it\theta(z)}{h_1}(z) \\
0 & 0 \\
 \end{array}
 \right)$,~~
 ${\omega_+^{a}}=\left(
\begin{array}{cc}
0 & {\delta}{_+^{2}}(z)e^{-2it\theta(z)}[\rho(z)]{h_2}(z) \\
0 & 0 \\
 \end{array}
 \right),$\\
$~~~~~~~~~~~~~~~~~~~~{\omega_-^{o}}=\left(
\begin{array}{cc}
0 & 0 \\
{\delta}{_-^{-2}}(z)e^{2it\theta(z)}\overline{{h_1}(\overline{z})} & 0 \\
 \end{array}
 \right)$,~~
 ${\omega_-^{a}}=\left(
\begin{array}{cc}
0 & 0 \\
{\delta}{_-^{-2}}(z)e^{2it\theta(z)}[\overline{\rho(\overline{z})}]\overline{{h_2}(\overline{z})} & 0 \\
 \end{array}
 \right)$.\\
Then we define the oriented contour $\Sigma^{(2)}$ as $\Sigma^{(2)}=\mathbb{R}\cup L \cup \overline{L}$, as shown in Fig. 3. By extending the jump matrix to the steep descend line $L$ and $\overline{L}$, $({b_-^{a}})^{-1}$ can be analytically extend to $\overline{L}$, ${b_+^{a}}$ can be analytically extend to $L$, $({b_-^{o}})^{-1}{b_+^{o}}$ do not have the property of analytic continuation but decays fast about time on $\mathbb{R}$. We make a transformation
\begin{equation}
m^{(2)}=m^{(1)}\phi,
\end{equation}
where
\begin{equation}
\phi=\begin{cases}
I,~~~~~~~~~z\in {\Omega_1}\cup{\Omega_2},\\
({b_-^{a}})^{-1},~~z\in {\Omega_3}\cup{\Omega_4}\cup{\Omega_5},\\
({b_+^{a}})^{-1},~~z\in {\Omega_6}\cup{\Omega_7}\cup{\Omega_8}.\\
\end{cases}
\end{equation}
Then Riemann-Hilbert problem in ${\Sigma^{(1)}}$ turns into a new Riemann-Hilbert problem in ${\Sigma^{(2)}}$, and $m^{(2)}$ satisfies
\begin{equation}
\begin{aligned}
&\bullet m^{(2)}(z,t,x) ~is ~analytic in~ \mathbb{C} \backslash{\Sigma}^{(2)},\\
&\bullet {m_+^{(2)}}(z,t,x) = {m_-^{(2)}}(z,t,x)v^{(2)}(z,t,x),~~z \in {\Sigma}^{(2)},\\
&\bullet {m_+^{(2)}}(z,t,x)\longrightarrow I,~~ as ~~z\rightarrow \infty,\\
\end{aligned}
\end{equation}
where the jump matrix is
\begin{equation}
v^{(2)}(x,t,z)=\begin{cases}
&({b_-^{o}})^{-1}{b_+^{o}},~~~~~z\in \mathbb{R},\\
&~~~~~{b_+^{a}},~~~~~~~~~z\in \overline{L},\\
&~~~({b_-^{a}}),~~~~~~~~~z\in L.\\
\end{cases}
\end{equation}
The solution of the Eq.(1.8) can be presented as
\begin{equation}
u(x,t)=2i\lim \limits_{z\to \infty}{\left(zm^{(2)}(z;x,t)\right)_{12}}.
\end{equation}
Firstly, in the case of $z \in {\Omega_6}$, considering the bound of ${\delta}(z)$ in Eq.(3.19), the definition of ${h_2}(z)$, and $\left[\rho(z)\right]$ in (3.32) and (3.33), we have
\begin{equation}
\begin{aligned}
&\left|\delta^{2}(z)e^{-2it\theta(z)}{h_2}(z)\right|\leq c \left|\gamma(z)\right|\cdot\left|e^{-itRe(i\theta(z))}\right|\cdot\left|\int_{-\infty}^{t}e^{i(s-t)\theta(z)}\hat{\left(\frac{h}{\gamma}\right)}(s)\overline{d}s\right|\\
&~~~~~~~~~~~~~~~~~~~~~~~~~~\leq \frac{\left|z-{z_0}\right|^{n}}{\left|z-i\right|^{n+2}}\cdot\int_{-\infty}^{t}\left|\hat{\left(\frac{h}{\gamma}\right)}(s)\right|\overline{d}s\leq\frac{c}{\left|z-i\right|^{2}},\\
\end{aligned}
\end{equation}
\begin{equation}
\left|\delta^{2}e^{-2it\theta(z)}\left[\rho(z)\right]\right|\leq \frac{c|\sum_{j=0}^{k}{\mu_i}(z-{z_0})^{j}|}{|z-i|^{k+5}}\leq \frac{c}{|z-i|^{5}}.
\end{equation}

\begin{figure}
  \centering
  \includegraphics[width=10cm]{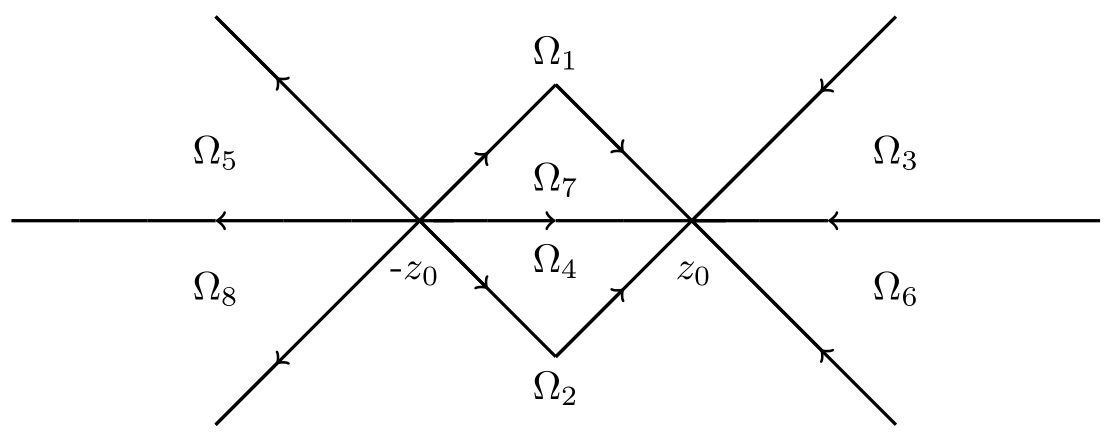}
  \caption{The oriented contour ${\Sigma^{(2)}}$.}
\end{figure}

Set
\begin{equation}
{v^{(2)}}(x,t,z)={\left({b_-^{(2)}}\right)}^{-1}{b_+^{(2)}},
\end{equation}
define
\begin{equation}
{\omega}{_\pm^{(2)}}=\pm\left({b_\pm^{(2)}}-I\right),~~{\omega^{(2)}}={\omega}{_+^{(2)}}+{\omega}{_-^{(2)}},
\end{equation}
and Cauchy operator
\begin{equation}
\left({C_\pm f}\right)(z)=\frac{1}{2\pi i}\int_{\Sigma^{(2)}} \frac{f(\xi)}{\xi-{z_\pm}}d\xi,~~z \in{\Sigma^{(2)}},~~f \in{\mathscr{L}^{2}}({\Sigma^{(2)}}).
\end{equation}
We note that Cauchy operator ${C_\pm}$ is a bounded operator from ${\mathscr{L}^{2}}({\Sigma^{(2)}})$ to ${\mathscr{L}^{2}}({\Sigma^{(2)}})$ and ${C_+}+{C_-}=1$. \\
Define
\begin{equation}
{C_{\omega^{(2)}}}f={C_+}(f{\omega_-^{(2)}})+{C_-}(f{\omega_+^{(2)}}),
\end{equation}
let ${\mu^{(2)}}={\mu^{(2)}}(z;x,t)\in {\mathscr{L}^{2}}({\Sigma^{(2)}})+{\mathscr{L}^{\infty}}({\Sigma^{(2)}})$ be the solution of the following basic inverse equation
\begin{equation}
{\mu^{(2)}}=I+{C_{\omega^{2}}}{\mu^{(2)}}.
\end{equation}
Then
\begin{equation}
m^{(2)}(z;x,t)=I+\frac{1}{2\pi i}\int_{\Sigma^{(2)}}\frac{\mu^{(2)}(\xi;x,t)\omega^{(2)}(\xi;x,t)}{\xi-z}d\xi
\end{equation}
is the solution of the Riemann-Hilbert problem (3.63). So the solution of Eq.(1.8) is
\begin{equation}
\begin{aligned}
&u(x,t)=2i\lim \limits_{z\to \infty}{\left(zm^{(2)}(z;x,t)\right)_{12}}\\
&~~~~~~~~=2i\lim \limits_{z\to \infty}{\left(z\left(I+{\frac{1}{2\pi i} }\int_{\Sigma^{(2)}}\frac{\mu^{(2)}(\xi;x,t)\omega^{(2)}(\xi;x,t)}{\xi-z}d\xi\right)\right)}{_{12}}\\
&~~~~~~~~=\lim \limits_{z\to \infty}z{\left(\int_{\Sigma^{(2)}}\frac{\mu^{(2)}(\xi;x,t)\omega^{(2)}(\xi;x,t)}{\xi-z}\frac{d\xi}{\pi}\right)}{_{12}}\\
&~~~~~~~~=-{\left(\int_{\Sigma^{(2)}}\frac{\mu^{(2)}(\xi;x,t)\omega^{(2)}(\xi;x,t)}{\pi}d\xi\right)}{_{12}}\\
&~~~~~~~~=-{\left(\int_{\Sigma^{(2)}}\left((I-{C_{\omega^{(2)}}})^{-1}\right)(\xi;x,t)\omega^{(2)}(\xi;x,t)\frac{d\xi}{\pi}\right)}{_{12}}.\\
\end{aligned}
\end{equation}
\subsection{ The third transformation}
In this subsection, the Riemann-Hilbert problem on the contour ${\Sigma^{(2)}}$ is converted to Riemann-Hilbert problem on the contour ${\Sigma^{(3)}}$ = ${\Sigma^{(2)}}\backslash (\mathbb{R}\cup {L_\varepsilon} \cup \overline{{L_\varepsilon}})$. We figure out the estimates of the errors between the two Riemann-Hilbert problems.
Let ${\omega^{e}}={\omega^{a}}+{\omega^{b}}+{\omega^{c}}$,
${\omega^{a}}={\omega^{(2)}}{|_\mathbb{R}}$ is supported on $\mathbb{R}$ and composed of terms of type ${h_1}(z)$, $\overline{{h_1}(\overline{z})}$. ${\omega^{b}}$ is supported on $L\cup \overline{L}$ and composed of the contribution to ${\omega^{(2)}}$ from terms of type ${h_2}(z)$, $\overline{{h_2}(\overline{z})}$.
${\omega^{c}}$ is supported on ${L_\varepsilon}\cup \overline{{L_\varepsilon}}$ and composed of the contribution to ${\omega^{(2)}}$ from terms of type $R(z)$ ,$\overline{R(\overline{z})}$.
That is
\begin{equation}
\begin{aligned}
&{\omega^{a}}=\begin{cases}
{\omega^{(2)}},~z \in \mathbb{R},\\
0,~~~~~ otherwise,\\
\end{cases}\\
&{\omega^{b}}=\begin{cases}
\left(
\begin{array}{cc}
0 & {\delta}{_+^{2}}e^{-2it\theta(z)}{h_2}(z) \\
0 & 0 \\
 \end{array}
 \right),~z \in L,\\
\left(
\begin{array}{cc}
0 & 0 \\
{\delta}{_-^{-2}}e^{2it\theta(z)}\overline{{h_2}(\overline{z})} & 0 \\
 \end{array}
 \right) ,~z \in \overline{L},\\
 0,~~~~~~~~~~~~~~~~~~~~~~~~~~~~~~~~~~otherwise
\end{cases}\\
&{\omega^{c}}=\begin{cases}
\left(
\begin{array}{cc}
0 & {\delta}{_+^{2}}e^{-2it\theta(z)}R(z) \\
0 & 0 \\
 \end{array}
 \right),~z \in {L_\varepsilon},\\
 \left(
\begin{array}{cc}
0 & 0 \\
{\delta}{_-^{-2}}e^{2it\theta(z)}\overline{R(\overline{z})} & 0 \\
 \end{array}
 \right),~~~z \in \overline{{L_\varepsilon}},\\
 0,~~~~~~~~~~~~~~~~~~~~~~~~~~~~~~~~~~otherwise.\\
 \end{cases}
\end{aligned}
\end{equation}
By defining ${\omega^{(3)}}$ as
\begin{equation}
{\omega^{(2)}}={\omega^{(3)}}+{\omega^{(e)}},
\end{equation}
we notice that ${\omega^{(3)}}=0$ on ${\Sigma^{(2)}}\backslash {\Sigma^{(3)}}$, which leads to the following estimates.\\
\textbf{Theorem 3.2}  For arbitrary positive integer $l$, as $t \rightarrow \infty$, we have the estimates
\begin{equation}
{\| \omega^{a} \|_{{\mathscr{L}^{\infty}(\mathbb{R})}\cap {\mathscr{L}^{2}(\mathbb{R})}\cap {\mathscr{L}^{1}(\mathbb{R})}}}\leq ct^{-l},
\end{equation}
\begin{equation}
{\| \omega^{b} \|_{{\mathscr{L}^{\infty}(L\cup \overline{L})}\cap {\mathscr{L}^{2}(L\cup \overline{L})}\cap {\mathscr{L}^{1}(L\cup \overline{L})}}}\leq ct^{-l},
\end{equation}
\begin{equation}
{\| \omega^{(3)}\|_{\mathscr{L}^{2}(\Sigma^{(2)})}}\leq \tau^{-\frac{1}{4}},~~{\| \omega^{(3)}\|_{\mathscr{L}^{1}(\Sigma^{(2)})}}\leq \tau^{-\frac{1}{2}}.
\end{equation}
\textbf{Proof.} Firstly, we consider the case of ${\omega^{a}}$. It is necessary to compute $|{\omega^{a}}|$ if we want to give the estimates for ${\| \omega^{a} \|_{\mathscr{L}^{p}(\mathbb{R})}}$. Notice that the bound of $\delta(z)$  in (3.19), we have
\begin{equation}
\begin{aligned}
&|\omega^{a}|=\left(\left|e^{-2it\theta(z)}\delta^{2}(z){h_1}(z)\right|^{2}+\left|e^{2it\theta(z)}\delta^{-2}(z)\overline{{h_1}(\overline{z})}\right|^{2}\right)^{\frac{1}{2}}\\
&~~~~~\leq \left|e^{-2it\theta(z)}\delta^{2}(z){h_1}(z)\right|+\left|e^{2it\theta(z)}\delta^{-2}(z)\overline{{h_1}(\overline{z})}\right|\\
&~~~~~\lesssim \left|e^{-2it\theta(z)}{h_1}(z)\right|+\left|e^{2it\theta(z)}\overline{{h_1}(\overline{z})}\right|\\
&~~~~~\lesssim \frac{1}{(|z|^{2}-1)t^{l}}.\\
\end{aligned}
\end{equation}
Similarly, we can prove that (3.77) also holds by a simple calculation. Next, we prove estimate (3.78). Note the definition of (3.35), we have
\begin{equation}
R(z)=\frac{\left|\sum_{j=0}^{m}{\mu_j}(z-{z_0})^{j}\right|}{\left|(z-i)^{m+5}\right|}\lesssim \frac{1}{1+|z|^{5}}.
\end{equation}
Due to $Re(i\theta(z))\geq 3\alpha {z_0}u^{2}$ on $L:{z={z_0}+{z_0}\beta e^{\frac{3\pi i}{4}} (-\infty<\beta\leq \infty)}$, we find that
\begin{equation}
\left|e^{-2it\theta(z)}\delta^{2}(z)R(z)\right|\lesssim e^{-6\alpha \varepsilon^{2}t{z_0}^{3}}\frac{1}{1+|z|^{5}}.\\
\end{equation}
From Proposition 2.23 and Corollary 2.25 in [17], we note that operator $\left(1-{C_{\omega^{(3)}}}\right)^{-1}$:$\mathscr{L}^{2}(\Sigma^{(2)})\rightarrow \mathscr{L}^{2}(\Sigma^{(2)})$ exists and is uniformly bounded. Moreover,
\begin{equation}
{\|\left(1-{C_{\omega^{(3)}}}\right)^{-1}\|}{_{\mathscr{L}^{2}(\Sigma^{(2)})}}\lesssim 1,~~{\|\left(1-{C_{\omega^{(2)}}}\right)^{-1}\|}{_{\mathscr{L}^{2}(\Sigma^{(2)})}}\lesssim 1.
\end{equation}

\begin{figure}
  \centering
  \includegraphics[width=7cm]{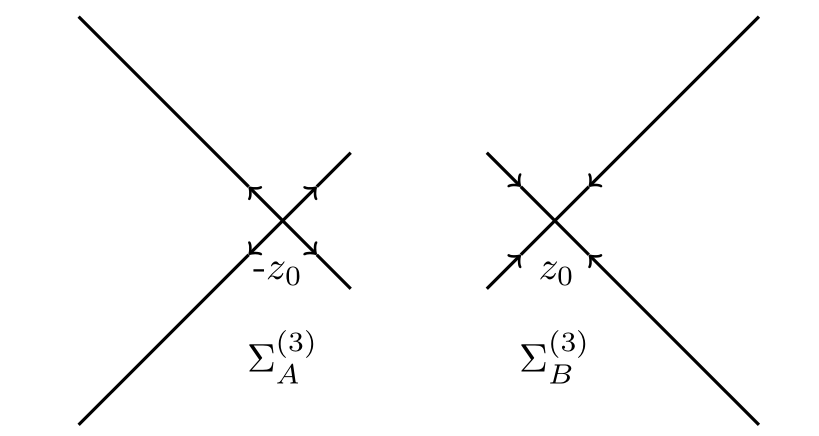}
  \caption{The oriented contour ${\Sigma^{(3)}}$.}
\end{figure}

\textbf{Theorem 3.3.} As $t\rightarrow \infty$, \\
\begin{equation}
\int_{\Sigma^{(2)}}\left((1-{C_{\omega^{(2)}}})^{-1}I\right)(\xi)\omega^{(2)}(\xi)d\xi=\int_{\Sigma^{(2)}}\left((1-{C_{\omega^{(3)}}})^{-1}I\right)(\xi)\omega^{(3)}(\xi)d\xi+o(t^{-l}).
\end{equation}
\textbf{Proof.} It is easy to see that
\begin{equation}
\left(1-{C_{\omega^{(2)}}}\right)^{-1}-\left(1-{C_{\omega^{(3)}}}\right)^{-1}=\left(1-{C_{\omega^{(3)}}}\right)^{-1}\left({C_{\omega^{(2)}}}-{C_{\omega^{(3)}}}\right)\left(1-{C_{\omega^{(2)}}}\right)^{-1},
\end{equation}
then
\begin{equation}
\begin{aligned}
&~~~~\left(1-{C_{\omega^{(2)}}}\right)^{-1}I\omega^{(2)}-\left(1-{C_{\omega^{(3)}}}\right)^{-1}I\omega^{(3)}\\
&=[\left(1-{C_{\omega^{(2)}}}\right)^{-1}-\left(1-{C_{\omega^{(3)}}}\right)^{-1}]I\omega^{(2)}+\left(1-{C_{\omega^{(3)}}}\right)^{-1}I(\omega^{(2)}-\omega^{(3)})\\
&=\left(1-{C_{\omega^{(3)}}}\right)^{-1}\left({C_{\omega^{(2)}}}-{C_{\omega^{(3)}}}\right)\left(1-{C_{\omega^{(2)}}}\right)^{-1}I\omega^{(2)}+\left(1-{C_{\omega^{(3)}}}\right)^{-1}I\omega^{e}\\
&=\left(1-{C_{\omega^{(3)}}}\right)^{-1}{C_{\omega^{e}}}\left(I+(1-{C_{\omega^{(3)}}})^{-1}{C_{\omega^{(2)}}}\right)I\omega^{(2)}+\left(I+(1-{C_{\omega^{(3)}}})^{-1}{C_{\omega^{(3)}}}\right)I\omega^{e}\\
&=\left(1-{C_{\omega^{(3)}}}\right)^{-1}{C_{\omega^{e}}}\omega+\left(1-{C_{\omega^{(3)}}}\right)^{-1}{C_{\omega^{e}}}\left(1-{C_{\omega^{(2)}}}\right)^{-1}{C_{\omega^{(2)}}}I{\omega^{(2)}}+
\omega^{e}+\left(1-{C_{\omega^{(3)}}}\right)^{-1}{C_{\omega^{(3)}}})I\omega^{e}.\\
\end{aligned}
\end{equation}
From theorem 3.1 and Eq(3.83), we have
\begin{equation}
\begin{aligned}
&{\|\omega^{e}\|}{_{\mathscr{L}^{1}(\Sigma^{2})}}\leq {\|\omega^{a}\|}{_{\mathscr{L}^{1}(\hat{\Gamma})}}+{\|\omega^{b}\|}{_{\mathscr{L}^{1}(L\cup\overline{L})}}+{\|\omega^{c}\|}{_{\mathscr{L}^{1}({L_\varepsilon}\cup\overline{{L_\varepsilon}})}}\\
&~~~~~~~~~~~~~\leq ct^{-l}+ct^{-l}+c{z_0}\tau^{-l}\leq c{z_0}\tau^{-l},~~\left(t^{l}\geq \frac{c\tau^{l}}{{z_0}}\right).\\
\end{aligned}
\end{equation}
By direct calculation, we get
\begin{equation}
\begin{aligned}
&~~~~{\|(1-{C_{\omega^{(3)}}})^{-1}({C_{\omega^{e}}I)}\omega^{(2)}\|}{_{\mathscr{L}^{1}(\Sigma^{(2)})}}\\
&\leq{\|(1-{C_{\omega^{(3)}}})^{-1}({C_{\omega^{e}}I)}\|}{_{{\mathscr{L}^{2}}(\Sigma^{(2)})}}{\|\omega^{(2)}\|}{_{\mathscr{L}^{2}(\Sigma^{(2)})}}\\
&\leq{\|(1-{C_{\omega^{(3)}}})^{-1}\|}{_{\mathscr{L}^{2}(\Sigma^{(2)})}}{\|{C_{\omega^{e}}I}\|}{_{\mathscr{L}^{2}(\Sigma^{(2)})}}+
\left({\|\omega^{a}\|}{_{\mathscr{L}^{2}(R)}}+{\|\omega^{b}\|}{_{\mathscr{L}^{1}(L\cup\overline{L})}}+{\|\omega^{(3)}\|}{_{\mathscr{L}^{2}(\Sigma^{(2)})}}\right)\\
&\lesssim{\|\omega^{e}\|}{_{\mathscr{L}^{2}(\Sigma^{(2)})}}\left(t^{-l}+t^{-l}+t^{-\frac{1}{4}}\right),\\
\end{aligned}
\end{equation}
and
\begin{equation}
\begin{aligned}
&~~~~{\|(1-{C_{\omega^{(3)}}})^{-1}({C_{\omega^{(3)}}}I){\omega^{e}}\|}{_{\mathscr{L}^{1}(\Sigma^{(2)})}}\\
&\leq{\|(1-{C_{\omega^{(3)}}})^{-1}({C_{\omega^{(3)}}}I)\|}{_{\mathscr{L}^{2}(\Sigma^{(2)})}}{\|\omega^{e}\|}{_{\mathscr{L}^{2}(\Sigma^{(2)})}}\\
&\leq{\|(1-{C_{\omega^{(3)}}})^{-1}\|}{_{\mathscr{L}^{2}(\Sigma^{(2)})}}{\|{C_{\omega^{(3)}}}I\|}{_{\mathscr{L}^{2}(\Sigma^{(2)})}}{\|\omega^{e}\|}{_{\mathscr{L}^{2}(\Sigma^{(2)})}}\\
&\lesssim{\|\omega^{(3)}\|}{_{\mathscr{L}^{2}(\Sigma^{(2)})}}\left({\|\omega^{a}\|}{_{\mathscr{L}^{2}(R)}}+{\|\omega^{b}\|}{_{\mathscr{L}^{1}(L\cup\overline{L})}}\right)\lesssim t^{-l-\frac{1}{4}},\\
\end{aligned}
\end{equation}
therefore,
\begin{equation}
\begin{aligned}
&~~~~{\|(1-{C_{\omega^{(3)}}})^{-1}{C_{\omega^{e}}}(1-{C_{\omega^{(2)}}})^{-1}{C_{\omega^{(2)}}}I{\omega^{(2)}}\|}{_{\mathscr{L}^{1}(\Sigma^{(2)})}}\\
&\leq{\|(1-{C_{\omega^{(3)}}})^{-1}{C_{\omega^{e}}}(1-{C_{\omega^{(2)}}})^{-1}{C_{\omega^{(2)}}}I\|}{_{\mathscr{L}^{2}(\Sigma^{(2)})}}{\|{\omega^{(2)}}\|}{_{\mathscr{L}^{2}(\Sigma^{(2)})}}\\
&\leq{\|(1-{C_{\omega^{(3)}}})^{-1}\|}{_{\mathscr{L}^{2}(\Sigma^{(2)})}}{\|(1-{C_{\omega^{(2)}}})^{-1}\|}{_{\mathscr{L}^{2}(\Sigma^{(2)})}}
{\|{C_{\omega^{e}}}\|}{_{\mathscr{L}^{2}(\Sigma^{(2)})}}{\|{C_{\omega^{(2)}}}I\|}{_{\mathscr{L}^{2}(\Sigma^{(2)})}}{\|{\omega^{(2)}}\|}{_{\mathscr{L}^{2}(\Sigma^{(2)})}}\\
&\leq{\|\omega^{e}\|}{_{\mathscr{L}^{\infty}(\Sigma^{(2)})}}{\|\omega^{(2)}\|}{_{\mathscr{L}^{2}(\Sigma^{(2)})}^2}\lesssim t^{-l-\frac{1}{2}}.\\
\end{aligned}
\end{equation}
For ${z_0}<M$, substituting the above estimates into (3.86), we obtain theorem 3.3.

\textbf{Notice} As $z \in \Sigma^{(2)} \backslash\Sigma^{(3)}$, $\omega^{(3)}(z)=0$, let $C_{\omega^{(3)}}|_{\mathscr{L}^2(\Sigma^{(3)})}$ denote the restriction of $C_{\omega^{(3)}}$ to $\mathscr{L}^2(\Sigma^{(3)})$. For simplicity, we write $C_{\omega^{(3)}}|_{\mathscr{L}^2(\Sigma^{(3)})}$ as $C_{\omega^{(3)}}$. Then
\begin{equation}
\int_{\Sigma^{(2)}}\left((1-C_{\omega^{(3)}})^{-1}I\right)(\xi)\omega^{(3)}(\xi)d\xi=\int_{\Sigma^{(3)}}\left((1-C_{\omega^{(3)}})^{-1}I\right)(\xi)\omega^{(3)}(\xi)d\xi.
\end{equation}
\textbf{Theorem 3.4} As $t\rightarrow \infty$, the solution $u(x,t)$ for the Cauchy problem of Eq.(1.8) has the asymptotic estimate
\begin{equation}
u(x,t)=-\frac{1}{\pi}{\left(\int_{\Sigma^{(3)}}((1-{C_{\omega^{(3)}}})^{-1})(\xi)\omega^{(3)}(\xi)d\xi\right)}{_{12}}+o\left(t^{-l}\right).
\end{equation}
\textbf{Proof.} A direct consequence of (3.74) and (3.84).\\
Set
\begin{equation}
m^{(3)}=I+\int_{\Sigma^{(3)}}\frac{\mu^{(3)}\omega^{(3)}(\xi,x,t)}{\xi-z}\frac{d\xi}{2\pi i},
\end{equation}
where $\mu^{(3)}=(1-{C_{\omega^{(3)}}})^{-1}I$, the solution (3.92) is equivalent to
\begin{equation}
u(x,t)=2i\lim \limits_{z\to \infty}{\left(zm^{(3)}(z,x,t)\right)}{_{12}}+o\left(t^{-l}\right).
\end{equation}
Note that $m^{(2)}(z,x,t)$ is the solution of the Riemann-Hilbert problem (3.62), we can construct
the following Riemann-Hilbert problem
\begin{equation}
\begin{cases}
{m_+^{(3)}}(z,x,t)={m_-^{(3)}}(z,x,t)v^{(3)}(z,x,t),~~z\in \Sigma^{(3)}\\
m^{(3)}(z,x,t)\to I,~~z\to\infty,\\
\end{cases}
\end{equation}
where\\
$~~~~~~~~~~~~~~~~~~~~~~~~v^{(3)}=\left({b_-^{(3)}}\right)^{-1}{b_+^{(3)}}=\left(I-{\omega_-^{(3)}}\right)^{-1}\left(I+{\omega_+^{(3)}}\right)$,\\
$~~~~~~~~~~~~~~~~~~~~~~~~{\omega^{(3)}}={\omega_+^{(3)}}+{\omega_-^{(3)}}$,\\
$~~~~~~~~~~~~~~~~~~~~~~~~{\omega_+^{(3)}}=\begin{cases}
\left(
\begin{array}{cc}
0 & {\delta}{_+^{2}}e^{-2it\theta(z)}\rho(z) \\
0 & 0 \\
 \end{array}
 \right),~z\in L,\\
 {0_{2\times2}},~~~~~~~~~~~~~~~~~~~~~~~~~~~~z\in \overline{L},\\
\end{cases}$\\
$~~~~~~~~~~~~~~~~~~~~~~~~{\omega_-^{(3)}}=\begin{cases}
{0_{2\times2}},~~~~~~~~~~~~~~~~~~~~~~~~~~~~z\in L,\\
\left(
\begin{array}{cc}
0 & 0 \\
{\delta}{_-^{-2}}e^{2it\theta(z)}\overline{\rho(\overline{z})} & 0 \\
 \end{array}
 \right),~~~z\in \overline{L}.\\
\end{cases}$\\
\subsection{Noninteraction of disconnected contour components}

Let the contour ${\Sigma^{(3)}}={\Sigma_A^{(3)}}\cup{\Sigma_B^{(3)}}$ and ${\omega_\pm^{(3)}}={\omega_{A\pm}^{(3)}}+{\omega_{B\pm}^{(3)}}$, where ${\omega_{{A\pm}^{(3)}}}(z)=0$ for $z\in{\Sigma_B^{(3)}}$. The operator ${C}{_{{\omega_{A}^{(3)}}}}$ and ${C}{_{{\omega_{B}^{(3)}}}}$: $\mathscr{L}^{2}(\Sigma^{(3)})+\mathscr{L}^{\infty}(\Sigma^{(3)})\to \mathscr{L}^{2}(\Sigma^{(3)})$ are defined in definition (3.71).\\
\textbf{Lemma 3.5}\\
$~~~~~~~~~~~~{\|{C}{_{{\omega_{B}^{(3)}}}}{C}{_{{\omega_{A}^{(3)}}}}\|}{_{\mathscr{L}^{2}(\Sigma^{(3)})}}={\|{C}{_{{\omega_{A}^{(3)}}}}
{C}{_{{\omega_{B}^{(3)}}}}\|}{_{\mathscr{L}^{2}(\Sigma^{(3)})}}\lesssim c({z_0})\tau^{-\frac{1}{2}}$,\\
$~~~~~~~~~~~~{\|{C}{_{{\omega_{B}^{(3)}}}}{C}{_{{\omega_{A}^{(3)}}}}\|}{_{{\mathscr{L}^{\infty}(\Sigma^{(3)})}\to {\mathscr{L}^{2}(\Sigma^{(3)})}}}$,~~${\|{C}{_{{\omega_{A}^{(3)}}}}{C}{_{{\omega_{B}^{(3)}}}}\|}{_{{\mathscr{L}^{\infty}(\Sigma^{(3)})}\to {\mathscr{L}^{2}(\Sigma^{(3)})}}}\lesssim c({z_0})\tau^{-\frac{3}{4}}$.\\
\textbf{Proof.}  See Lemma 3.5 in [17].\\
\textbf{Theorem 3.6}\\
As $\tau \to \infty$, we have
\begin{equation}
\begin{aligned}
&~~~~\int_{\Sigma^{(3)}}\left((1-{C_{\omega^{(3)}}})^{-1}I\right)(\xi)\omega^{(3)}(\xi)d\xi\\
&=\int_{\Sigma{_A^{(3)}}}\left(1-{C{_{{\omega{_A^{(3)}}}}}}\right)^{-1}I(\xi){\omega{_A^{(3)}}}(\xi)d\xi+
\int_{\Sigma{_B^{(3)}}}\left(1-{C{_{{\omega{_B^{(3)}}}}}}\right)^{-1}I(\xi){\omega{_B^{(3)}}}(\xi)d\xi+o\left(\frac{c({z_0})}{\tau}\right).\\
\end{aligned}
\end{equation}
\textbf{Proof.} From identity
\begin{equation}
\begin{aligned}
&\left(1-{C{_{{\omega{_A^{(3)}}}}}}-{C{_{{\omega{_B^{(3)}}}}}}\right)\left(1+{C{_{{\omega{_A^{(3)}}}}}}(1-{C{_{{\omega{_A^{(3)}}}}}})^{-1}
+{C{_{{\omega{_B^{(3)}}}}}}(1-{C{_{{\omega{_B^{(3)}}}}}})^{-1}\right)\\
&=1-{C{_{{\omega{_B^{(3)}}}}}}{C{_{{\omega{_A^{(3)}}}}}}\left(1-{C{_{{\omega{_A^{(3)}}}}}}\right)^{-1}-
{C{_{{\omega{_A^{(3)}}}}}}{C{_{{\omega{_B^{(3)}}}}}}\left(1-{C{_{{\omega{_B^{(3)}}}}}}\right)^{-1},
\end{aligned}
\end{equation}
we have
\begin{equation}
\begin{aligned}
&\left(1-{C_{\omega^{(3)}}}\right)^{-1}=1+{C{_{{\omega{_A^{(3)}}}}}}\left(1-{C{_{{\omega{_A^{(3)}}}}}}\right)^{-1}
{C{_{{\omega{_B^{(3)}}}}}}\left(1-{C{_{{\omega{_B^{(3)}}}}}}\right)^{-1}+\left[1+{C{_{{\omega{_A^{(3)}}}}}}\left(1-{C{_{{\omega{_A^{(3)}}}}}}\right)^{-1}
{C{_{{\omega{_B^{(3)}}}}}}\left(1-{C{_{{\omega{_B^{(3)}}}}}}\right)^{-1}\right]\\
&~~~~~~~~~~~~~~~~~~~~~
\cdot\left[1-{C{_{{\omega{_B^{(3)}}}}}}{C{_{{\omega{_A^{(3)}}}}}}\left(1-{C{_{{\omega{_A^{(3)}}}}}}\right)^{-1}
-{C{_{{\omega{_A^{(3)}}}}}}{C{_{{\omega{_B^{(3)}}}}}}\left(1-{C{_{{\omega{_B^{(3)}}}}}}\right)^{-1}\right]^{-1}\\
&~~~~~~~~~~~~~~~~~~~~~
\cdot\left[{C{_{{\omega{_B^{(3)}}}}}}{C{_{{\omega{_A^{(3)}}}}}}\left(1-{C{_{{\omega{_A^{(3)}}}}}}\right)^{-1}
+{C{_{{\omega{_A^{(3)}}}}}}{C{_{{\omega{_B^{(3)}}}}}}\left(1-{C{_{{\omega{_B^{(3)}}}}}}\right)^{-1}\right].\\
\end{aligned}
\end{equation}
By using theorem (3.2), Eq.(3.84) and Lemma 3.5, we obtain theorem (3.6). For convenience, we write the restriction ${C{_{{\omega{_A^{(3)}}}}}}|_{\mathscr{L}^{2}({\Sigma_A^{(3)}})}$ as ${C{_{{\omega{_A^{(3)}}}}}}$. Similar for ${C{_{{\omega{_B^{(3)}}}}}}$.\\
\textbf{Lemma3.7} As $\tau \to \infty$,\\
\begin{equation}
u(x,t)
=-\frac{1}{\pi}{\left(\int_{\Sigma_A^{(3)}}((1-{C_{\omega_A^{(3)}}})^{-1})(\xi){\omega_A^{(3)}}(\xi)d\xi\right)}{_{12}}
-\frac{1}{\pi}{\left(\int_{\Sigma_B^{(3)}}((1-{C_{\omega_B^{(3)}}})^{-1})(\xi){\omega_B^{(3)}}(\xi)d\xi\right)}{_{12}}
+o\left(\frac{c({z_0})}{\tau}\right).
\end{equation}
\subsection{Rescaling and further reduction of the RH problems}
Then we extend the contours ${\Sigma_A^{(3)}}$, ${\Sigma_B^{(3)}}$ to the contour\\
$~~~~~~~~~~~~~~~~{\hat{\Sigma}}{_A^{(3)}}:{z=-{z_0}+{z_0}\eta e^{\pm \frac{\pi i}{4}} (\eta\in \mathbb{R}})$,
${\hat{\Sigma}}{_B^{(3)}}:{z={z_0}+{z_0}\eta e^{\pm \frac{3\pi i}{4}} (\eta\in \mathbb{R}})$,\\
and define ${\hat{\omega}}{_A^{(3)}}$, ${\hat{\omega}}{_B^{(3)}}$ as
 \begin{equation}
 {\hat{\omega}}{_{A\pm}^{(3)}}=
 \begin{cases}
 {\omega_{A\pm}^{(3)}}(z), ~z\in {\Sigma_A^{(3)}},\\
 0,~~~~~~~~~z \in {\hat{\Sigma}_A^{(3)}}\backslash {\Sigma_A^{(3)}},\\
 \end{cases}
\end{equation}
\begin{equation}
 {\hat{\omega}}{_{B\pm}^{(3)}}=
 \begin{cases}
 {\omega_{B\pm}^{(3)}}(z), ~z\in {\Sigma_B^{(3)}},\\
 0,~~~~~~~~~z \in {\hat{\Sigma}_B^{(3)}}\backslash {\Sigma_B^{(3)}}.\\
 \end{cases}
\end{equation}

\begin{figure}
  \centering
  \includegraphics[width=10cm]{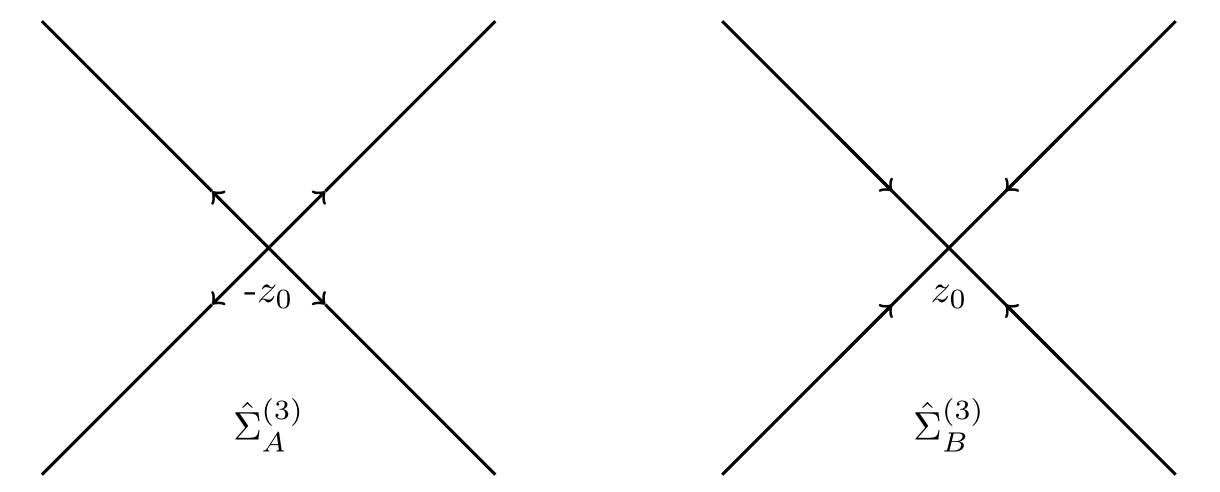}
  \caption{The oriented contour ${\hat{\Sigma}^{(3)}}$.}
\end{figure}

Let ${\Sigma_A}$ and ${\Sigma_B}$ denote the contour ${z={z_0}\eta e^{\pm \frac{\pi i}{4}} (\eta\in \mathbb{R}})$.\\

\begin{figure}
  \centering
  \includegraphics[width=10cm]{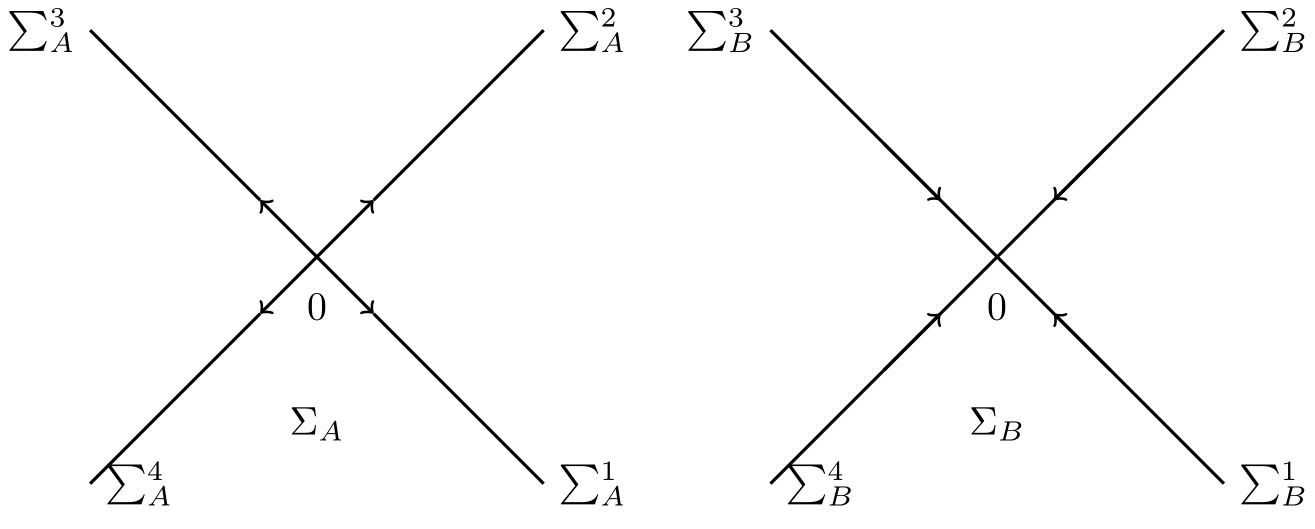}
  \caption{The oriented contour $\Sigma$.}
\end{figure}

Define the scaling operator as
\begin{equation}
\begin{aligned}
&{N_A}: \mathscr{L}^{2}\left({\hat{\Sigma}}{_{A}^{(3)}}\right) \to \mathscr{L}^{2}\left({\Sigma_A}\right),\\
&f(z)\to \left({N_A}f\right)(z)=f\left(\frac{z}{\sqrt{12\alpha t{z_0}}}-{z_0}\right),\\
\end{aligned}
\end{equation}

\begin{equation}
\begin{aligned}
&{N_B}: \mathscr{L}^{2}\left({\hat{\Sigma}}{_{B}^{(3)}}\right) \to \mathscr{L}^{2}({\Sigma_B}),\\
&f(z)\to ({N_B}f)(z)=f\left(\frac{z}{\sqrt{12\alpha t{z_0}}}+{z_0}\right).\\
\end{aligned}
\end{equation}
Take ${\omega_A}={N_A}{\hat{\omega}}{_{A\pm}^{(3)}}$, ${\omega_B}={N_B}{\hat{\omega}}{_{B\pm}^{(3)}}$,
and the bounded operator ${C_{\omega_A}}$ :$\mathscr{L}^{2}({\Sigma_A})(\mathscr{L}^{2}({\Sigma_B}))
\to \mathscr{L}^{2}\left({\Sigma_A}\right)\left(\mathscr{L}^{2}({\Sigma_B})\right)$. For a $2\times2$ matrix-valued function $f$, a direct calculation shows
\begin{equation}
N{C_{\hat{\omega}}{_{A}^{(3)}}}f=N\left({C_+}(f{\hat{\omega}}{_{A-}^{(3)}})+{C_-}(f{\hat{\omega}}{_{A+}^{(3)}})\right)
={C_{\hat{\omega}}{_{A}^{(3)}}}Nf,
\end{equation}
which means
\begin{equation}
{C_{\hat{\omega}}{_{A}^{(3)}}}={N_A}^{-1}{C_{\omega_A}}{N_A}.
\end{equation}
Similarly, we get
\begin{equation}
{C_{\hat{\omega}}{_{B}^{(3)}}}={N_B}^{-1}{C_{\omega_B}}{N_B}.
\end{equation}
From the definition of ${\omega_A}$, it is easy to see that
\begin{equation}
{\omega_A}={\omega_{A}+}=
\left(
\begin{array}{cc}
0 & {N_A}({\delta}{_+^{2}}(z)e^{-2it\theta(z)}\rho(z)) \\
0 & 0 \\
 \end{array}
 \right)
\end{equation}
on ${L_A}$, and
\begin{equation}
{\omega_A}={\omega_{A}-}=
\left(
\begin{array}{cc}
0 & 0 \\
{N_A}({\delta}{_-^{-2}}(z)e^{2it\theta(z)}\overline{\rho(\overline{z})}) & 0 \\
 \end{array}
 \right)
\end{equation}
on $\overline{{L_A}}$.
By using formulae (3.4) and (3.16), we obtain
\begin{equation}
{N_A}\left({\delta_A}(z) e^{-it\theta(z)}\right)={\delta_A^0}{\delta_A^1}(z),
\end{equation}
where\\
$~~~~~~~~~{\delta_A^0}=e^{-2i\alpha\tau}(12\alpha\tau)^{\frac{i\upsilon}{2}}e^{\chi(-{z_0})}$\\
and\\
$~~~~~~~~~{\delta_A^1}(z)=z^{-i\upsilon}\left(\frac{{z_0}}{\frac{z}{\sqrt{12\alpha t{z_0}}}-2{z_0}}\right)^{-i\upsilon}
e^{\frac{iz^{2}}{4}\left(1-z(108\alpha\tau)^{-\frac{1}{2}}\right)}e^{\chi\left(\frac{z}{\sqrt{12\alpha t{z_0}}}-{z_0}\right)-\chi\left(-{z_0}\right)}$.\\
In addition, through formulae (3.5) and (3.16), we get
\begin{equation}
{N_B}\left({\delta_B}(z) e^{-it\theta(z)}\right)={\delta_B^0}(z){\delta_B^1}(z),
\end{equation}
where\\
$~~~~~~~~~{\delta_B^0}=e^{2i\alpha\tau}(12\alpha\tau)^{-\frac{i\upsilon}{2}}e^{\chi({z_0})}$\\
and\\
$~~~~~~~~~{\delta_B^1}(z)=z^{i\upsilon}\left(\frac{{z_0}}{\frac{z}{\sqrt{12\alpha t{z_0}}}+2{z_0}}\right)^{i\upsilon}
e^{-\frac{iz^{2}}{4}\left(1+z(108\alpha\tau)^{-\frac{1}{2}}\right)}e^{\chi\left(\frac{z}{\sqrt{12\alpha t{z_0}}}+{z_0}\right)-\chi\left({z_0}\right)}$.\\
Notice that ${\delta_A^0}$ and ${\delta_B^0}$ are independent of $z$, and
\begin{equation}
\left|{\delta_A^0}\right|=1,~\left|{\delta_B^0}\right|=1.
\end{equation}
Here we construct ${\omega_0}$ on the contour ${\Sigma_A}$ and ${\Sigma_B}$ respectively. Set $v^{A0}=\left(I-{\omega_{A^0-}}\right)^{-1}\left(I+{\omega_{A^0+}}\right)$, where
\begin{equation}
{\omega_{A^0}}={\omega_{A^0+}}=
\begin{cases}
\left(
\begin{array}{cc}
0 & ({\delta_A^0})^{2}(-z)^{-2i\upsilon}e^{\frac{iz^{2}}{2}}\overline{r(-{z_0})} \\
0 & 0 \\
 \end{array}
 \right),z\in {\Sigma_A^4},\\
 \left(
\begin{array}{cc}
0 & ({\delta_A^0})^{2}(-z)^{-2i\upsilon}e^{\frac{iz^{2}}{2}}\frac{-\overline{r(-{z_0})}}{1-|r({z_0})|^{2}} \\
0 & 0 \\
 \end{array}
 \right),z\in {\Sigma_A^2},\\
\end{cases}
\end{equation}
\begin{equation}
{\omega_{A^0}}={\omega_{A^0-}}=
\begin{cases}
\left(
\begin{array}{cc}
0 & 0 \\
({\delta_A^0})^{-2}(-z)^{2i\upsilon}e^{-\frac{iz^{2}}{2}}\overline{r(-{z_0})} & 0 \\
 \end{array}
 \right),z\in {\Sigma_A^3},\\
 \left(
\begin{array}{cc}
0 & 0 \\
({\delta_A^0})^{-2}(-z)^{2i\upsilon}e^{-\frac{iz^{2}}{2}}\frac{r({z_0})}{1-|r({z_0})|^{2}} & 0 \\
 \end{array}
 \right),z\in {\Sigma_A^1},\\
\end{cases}
\end{equation}
\begin{equation}
{\delta_A^0}=e^{-2i\alpha\tau}(12\alpha\tau)^{\frac{i\upsilon}{2}}e^{\chi(-{z_0})}.
\end{equation}

Due to Lemma 3.35 in [17], we have\\
$~~~~~~~~~~~~~~~~~~~~~~~~~~~~~~~~~~~~~~~~~~~~{\|{\omega_A}-{\omega_A^0}\|}{_{{\mathscr{L}^{\infty}({\Sigma_A})}\cap {\mathscr{L}^{2}({\Sigma_A})}\cap {\mathscr{L}^{1}({\Sigma_A})}}}\lesssim t^{-\frac{1}{2}}\log t$.

As $t \to \infty$, we have
\begin{equation}
\begin{aligned}
&~~~~\int_{{{\Sigma}}{_A^{(3)}}}((1-{C_{{\omega}}{_{A}^{(3)}}})^{-1}I){{\omega}}{_{A}^{(3)}}(\xi)d\xi\\
&=\int_{{\hat{\Sigma}}{_A^{(3)}}}((1-{C_{\hat{\omega}}{_{A}^{(3)}}})^{-1}I){\hat{\omega}}{_{A}^{(3)}}(\xi)d\xi\\
&=\int_{{\hat{\Sigma}}{_A^{(3)}}}{N_A}^{-1}((1-{C_{\omega A}})^{-1}{N_A}I)(\xi){\hat{\omega}}{_{A}^{(3)}}(\xi)d\xi\\
&=\int_{{\hat{\Sigma}}{_A^{(3)}}}((1-{C_{\omega_A}})^{-1}I)\left((\xi+{z_0})\sqrt{12\alpha t}\right){N_A}{\hat{\omega}}{_{A}^{(3)}}\left((\xi+{z_0})\sqrt{12\alpha t}\right)d\xi\\
&=\frac{1}{\sqrt{12\alpha t}}\int_{{\Sigma_A}}\left((1-{C_{\omega A}})^{-1}I)\right)(\xi){\omega_A}(\xi)d\xi\\
&=\frac{1}{\sqrt{12\alpha t}}\int_{{\Sigma_A}}\left((1-{C_{\omega_A^0}})^{-1}I\right)(\xi){\omega_A^0}(\xi)d\xi+o\left(\frac{\log t}{t}\right).\\
\end{aligned}
\end{equation}
Similarly,
\begin{equation}
\begin{aligned}
&~~~~\int_{{{\Sigma}}{_B^{(3)}}}\left((1-{C_{{\omega}}{_{B}^{(3)}}})^{-1}I\right){{\omega}}{_{B}^{(3)}}(\xi)d\xi\\
&=\frac{1}{\sqrt{12\alpha t}}\int_{{\Sigma_B}}\left((1-{C_{\omega_B^0}})^{-1}I\right)(\xi){\omega_B^0}(\xi)d\xi+o\left(\frac{\log t}{t}\right).\\
\end{aligned}
\end{equation}
On the basis of above analysis, we have
\begin{equation}
\begin{aligned}
&u(x,t)=-{\left(\frac{1}{\sqrt{12\alpha t}}\int_{{\Sigma_A}}((1-{C_{\omega_A^0}})^{-1}I)(\xi){\omega_A^0}(\xi )\frac{d\xi}{\pi}\right)}{_{12}}\\
&~~~~~~~~~~~-{\left(\frac{1}{\sqrt{12\alpha t}}\int_{{\Sigma_B}}((1-{C_{\omega_B^0}})^{-1}I)(\xi){\omega_B^0}(\xi )\frac{d\xi}{\pi}\right)}{_{12}}\\
&~~~~~~~~~~~+o\left(\frac{\log t}{t}\right).\\
\end{aligned}
\end{equation}
For $z\in {\Sigma_A}$, set
\begin{equation}
{M^{A0}}(z,x,t)=I+\int_{\Sigma_A}\frac{\left((1-{C_{\omega_A^0}})^{-1}I\right)(\xi){\omega_A^0}(\xi)}{\xi-z}\frac{d\xi}{2\pi i}.
\end{equation}
Then ${M^{A0}}(z,x,t)$ is the solution of the following Riemann-Hilbert problem
\begin{equation}
\begin{cases}
{M_+^{A0}}(z,x,t)={M_-^{A0}}(z,x,t)v^{{A0}}(z,x,t),~~z\in {\Sigma_A},\\
{M^{A0}}(z,x,t)\to I,~~z\to \infty.
\end{cases}
\end{equation}
Since ${M^{A0}}(z)=I+\frac{{M_1^{A0}}}{z}+o\left(z^{-2}\right)$, $z\to \infty$, we have
${M_1^{A0}}=-\int_{{\Sigma_A}}\left((1-{C_{\omega_A^0}})^{-1}I\right){\omega_A^0}(\xi)\frac{d\xi}{2\pi i}$.\\
For $z\in {\Sigma_B}$, set
\begin{equation}
{M^{B0}}(z,x,t)=I+\int_{\Sigma_B}\frac{\left((1-{C_{\omega_B^0}})^{-1}I\right)(\xi){\omega_B^0}}{\xi-z}\frac{d\xi}{2\pi i}.
\end{equation}
then ${M^{B0}}(z,x,t)$ is the solution of the following Riemann-Hilbert problem
\begin{equation}
\begin{cases}
{M_+^{B0}}(z,x,t)={M_-^{B0}}(z,x,t)v^{{B0}}(z,x,t),~~z\in {\Sigma_B},\\
{M^{B0}}(z,x,t)\to I,~~z\to \infty.
\end{cases}
\end{equation}
Since ${M^{B0}}(z)=I+\frac{{M_1^{B0}}}{z}+o\left(z^{-2}\right)$, $z\to \infty$,
we have ${M_1^{B0}}=-\int_{{\Sigma_B}}\left((1-{C_{\omega_B^0}})^{-1}I\right){\omega_B^0}(\xi)\frac{d\xi}{2\pi i}$.
$v^{B0}=\left(I-{\omega_{B^0-}}\right)^{-1}\left(I+{\omega_{B^0+}}\right)$, where
\begin{equation}
{\omega_B^0}={\omega_{B^0+}}=
\begin{cases}
\left(
\begin{array}{cc}
0 & ({\delta_B^0})^{2}z^{2i\upsilon}e^{-\frac{iz^{2}}{2}}\overline{r({z_0})} \\
0 & 0 \\
 \end{array}
 \right),z\in {\Sigma_B^1},\\
 \left(
\begin{array}{cc}
0 & ({\delta_B^0})^{2}z^{2i\upsilon}e^{-\frac{iz^{2}}{2}}\frac{-\overline{r({z_0})}}{1-|r({z_0})|^{2}} \\
0 & 0 \\
 \end{array}
 \right),z\in {\Sigma_B^3},\\
\end{cases}
\end{equation}
\begin{equation}
{\omega_B^0}={\omega_{B^0-}}=
\begin{cases}
\left(
\begin{array}{cc}
0 & 0 \\
({\delta_B^0})^{-2}z^{-2i\upsilon}e^{\frac{iz^{2}}{2}}r({z_0}) & 0 \\
 \end{array}
 \right),z\in {\Sigma_B^2},\\
 \left(
\begin{array}{cc}
0 & 0 \\
({\delta_B^0})^{-2}z^{-2i\upsilon}e^{\frac{iz^{2}}{2}}\frac{r({z_0})}{1-|r({z_0})|^{2}} & 0 \\
 \end{array}
 \right),z\in {\Sigma_B^4},\\
\end{cases}
\end{equation}
\begin{equation}
{\delta_B^0}=e^{2i\alpha\tau}(12\alpha\tau)^{-\frac{i\upsilon}{2}}e^{\chi({z_0})}.
\end{equation}
By using (3.112)-(3.114) and (3.122)-(3.124), we have
\begin{equation}
v^{A0}(z)=\overline{(v^{B0})}(-\overline{z}).
\end{equation}
According to the uniqueness of the solution,
\begin{equation}
M^{A0}(z)=\overline{(M^{B0})}(-\overline{z}),
\end{equation}
and
\begin{equation}
{M_1^{A0}}=-\overline{({M_1^{B0}})}.
\end{equation}
Therefore,
\begin{equation}
u(x,t)=\frac{i}{\sqrt{3\alpha t{z_0}}}{\left({M_1^{B0}}-\overline{{M_1^{B0}}}\right)}{_{12}}.
\end{equation}

\subsection{ Solving the model problem}
In this subsection, our work is devoted to compute ${\left({M_1^{B0}}\right)}{_{12}}$ explicitly. Firstly, we introduce the transformation
$\psi(z)=H(z)z^{i\upsilon{\sigma_3}}e^{-\frac{iz^{2}}{4}{\sigma_3}}$,
$H(z)=({\delta_B^0})^{{-\sigma_3}}{M_B^0}(z)({\delta_B^0})^{{\sigma_3}}$,
then
\begin{equation}
\begin{aligned}
&{\psi_+}(z)=({\delta_B^0})^{{-\sigma_3}}{M_{B+}^0}(z)({\delta_B^0})^{{\sigma_3}}z^{i\upsilon{\sigma_3}}e^{-\frac{iz^{2}}{4}{\sigma_3}}\\
&~~~~~~~~=({\delta_B^0})^{{-\sigma_3}}{M_{B-}^0}(z)v^{B0}(z)({\delta_B^0})^{{\sigma_3}}z^{i\upsilon{\sigma_3}}e^{-\frac{iz^{2}}{4}{\sigma_3}}\\
&~~~~~~~~={\psi_-}(z)v({z_0}),\\
\end{aligned}
\end{equation}
where
$v({z_0})=z^{-i\upsilon{\hat{\sigma}}{_3}}e^{\frac{iz^{2}}{4}{\hat{\sigma}}{_3}}({\delta_B^0})^{{-\hat{\sigma}}{_3}}
v^{B0}$. Notice that the jump matrix is irrelevant to $z$ on each ray, we have
\begin{equation}
\frac{d{\psi_+}(z)}{dz}=\frac{{d{\psi_-}(z)}}{dz}v({z_0})
\end{equation}
and
\begin{equation}
\left(\frac{d{\psi_+}(z)}{dz}\right){\psi_+^{-1}(z)}=\frac{{d{\psi_-}(z)}}{dz}v({z_0})v^{-1}({z_0}){\psi_-^{-1}(z)}
=\frac{{d{\psi_-}(z)}}{dz}{\psi_-^{-1}(z)}.
\end{equation}
Therefore, $\frac{d\psi}{dz}\psi^{-1}(z)$ has no jump discontinuity along any of the ray. Moreover, from the relation between $\psi(z)$ and $H(z)$, we have
\begin{equation}
\begin{aligned}
&\frac{d\psi}{dz}\psi^{-1}(z)=\frac{dH(z)}{dz}H^{-1}(z)-\frac{i}{2}zH(z){\sigma_3}H^{-1}(z)\\
&~~~~~~~~~~~~~~=-\frac{i}{2}z{\sigma_3}-\frac{i}{2}({\delta_B^0})^{-{\hat{\sigma}}{_3}}\left[{M_1^{B0}},{\sigma_3}\right].\\
\end{aligned}
\end{equation}
It follows by the Liouville's theorem that
\begin{equation}
\frac{d\psi(z)}{dz}+\frac{i}{2}z{\sigma_3}\psi(z)=\beta \psi(z),
\end{equation}
where
$\beta=-\frac{i}{2}({\delta_B^0})^{-{\hat{\sigma}}{_3}}[{M_1^{B0}},{\sigma_3}]
=\left(
\begin{array}{cc}
0 & {\beta_{12}} \\
{\beta_{21}} & 0 \\
 \end{array}
 \right)$.
So we have
\begin{equation}
{M_1^{B0}}{_{12}}=-i({\delta_B^0})^{2}{\beta_{12}}.
\end{equation}
Set\\
$~~~~~~~~~~~~~~~~~~~~~~~~~~~~~~~~~~~~~~~~~~~~~~~~~~~~~~~\psi(z)=\left(
\begin{array}{cc}
{\psi_{11}}(z) & {\psi_{12}}(z) \\
{\psi_{21}}(z) & {\psi_{22}}(z) \\
 \end{array}
 \right)$.\\
From Eq.(3.133), we get
\begin{equation}
\frac{d^{2}{\psi_{11}}(z)}{dz^{2}}=\left(-\frac{i}{2}-\frac{z^{2}}{4}+{\beta_{12}}{\beta_{21}}\right){\psi_{11}}(z),
\end{equation}
\begin{equation}
{\beta_{12}}{\psi_{21}}(z)=\frac{d{\psi_{11}}(z)}{dz}+\frac{i}{2}z{\psi_{11}}(z),
\end{equation}
\begin{equation}
\frac{d^{2}{\psi_{22}}(z)}{dz^{2}}=\left(\frac{i}{2}-\frac{z^{2}}{4}+{\beta_{12}}{\beta_{21}}\right){\psi_{22}}(z),
\end{equation}
\begin{equation}
{\beta_{21}}{\psi_{12}}(z)=\frac{d{\psi_{22}}(z)}{dz}-\frac{i}{2}z{\psi_{22}}(z).
\end{equation}
We note that Weber's equation
\begin{equation}
\frac{d^{2}g(\xi)}{d{\xi^{2}}}+\left(a+\frac{1}{2}-\frac{\xi^{2}}{4}\right)g(\xi)=0
\end{equation}
has the solution
\begin{equation}
g(\xi)={c_1}{D_a}(\xi)+{c_2}{D_a}(-\xi),
\end{equation}
where ${D_a}(\cdot)$ denotes the standard parabolic-cylinder function and satisfies
\begin{equation}
\frac{d{D_a}(\xi)}{d\xi}+\frac{\xi}{2}{D_a}(\xi)-a{D_{a-1}}(\xi)=0,
\end{equation}
\begin{equation}
{D_a}(\pm\xi)=\frac{\Gamma(a+1)e^{\frac{i\pi a}{2}}}{\sqrt{2\pi}}{D_{-a-1}}\left(\pm i\xi\right)
+\frac{\Gamma(a+1)e^{-\frac{i\pi a}{2}}}{\sqrt{2\pi}}{D_{-a-1}}\left(\mp i\xi\right).
\end{equation}

According to [26], as $\xi\to \infty$, we have
\begin{equation}
{D_a}(\xi)=\begin{cases}
\xi^{a}e^{-\frac{\xi^{2}}{4}}\left(1+o(\xi^{-2})\right),~~~~~~~~~~~~~~~~~~~~~~~~~~~~~~~~~~~~~~~~~~~~~~|arg\xi|<\frac{3\pi}{4},\\
\xi^{a}e^{-\frac{\xi^{2}}{4}}\left(1+o(\xi^{-2})\right)-\frac{\sqrt{2\pi}}{\Gamma(-a)}e^{a\pi i+\frac{\xi^{2}}{4}}
\xi^{-a-1}\left(1+o(\xi^{-2})\right),~~\frac{\pi}{4}<arg\xi<\frac{5\pi}{4},\\
\xi^{a}e^{-\frac{\xi^{2}}{4}}\left(1+o(\xi^{-2})\right)-\frac{\sqrt{2\pi}}{\Gamma(-a)}e^{-a\pi i+\frac{\xi^{2}}{4}}
\xi^{-a-1}\left(1+o(\xi^{-2})\right),~~-\frac{5\pi}{4}<arg\xi<-\frac{\pi}{4},\\
\end{cases}
\end{equation}
where $\Gamma(\cdot)$ is the Gamma function.
Setting $a=i{\beta_{12}}{\beta_{21}}$, we have
\begin{equation}
{\psi_{11}}(z)={c_3}{D_a}\left(e^{-\frac{3\pi i}{4}}z\right)+{c_4}{D_a}\left(e^{\frac{\pi i}{4}}z\right)
\end{equation}
and
\begin{equation}
{\beta_{12}}{\psi_{22}}(z)={c_5}{D_{-a}}\left(e^{\frac{3\pi i}{4}}z\right)+{c_6}{D_{-a}}\left(e^{-\frac{\pi i}{4}}z\right).
\end{equation}
When $arg(z)\in \left(-\frac{\pi}{4},\frac{\pi}{4}\right)$ and $z\to \infty$, we have
${\psi_{11}}(z)\to z^{i\upsilon}e^{-\frac{iz^{2}}{4}}$, ${\psi_{22}}(z)\to z^{-i\upsilon}e^{\frac{iz^{2}}{4}}$. By using Eq.(3.143) and (3.144), we obtain
\begin{equation}
{c_3}=0,~{c_4}=e^{\frac{\pi\upsilon}{4}},~a=i\upsilon,~~{\psi_{11}}(z)=e^{\frac{\pi\upsilon}{4}}
{D_a}\left(e^{\frac{\pi i}{4}}z\right).
\end{equation}
Similarly, by using Eq.(3.143) and (3.145), we get
\begin{equation}
{c_5}=0,~{c_6}={\beta_{12}}e^{\frac{\pi\upsilon}{4}},~a=i\upsilon,~~{\beta_{12}}(z){\psi_{22}}(z)={\beta_{12}}(z)e^{\frac{\pi\upsilon}{4}}
{D_{-a}}\left(e^{-\frac{\pi i}{4}}z\right).
\end{equation}
Considering Eq.(3.133) and (3.136), we obtain
\begin{equation}
{\beta_{21}}{\psi_{21}}(z)=ae^{\frac{\pi(\upsilon+i)}{4}}{D_{a-1}}\left(e^{\frac{\pi i}{4}}z\right).
\end{equation}
When $arg(z)\in \left(-\frac{3\pi}{4},-\frac{\pi}{4}\right)$ and $z\to \infty$, we have
${\psi_{11}}(z)\to z^{i\upsilon}e^{-\frac{iz^{2}}{4}}$, ${\psi_{22}}(z)\to z^{-i\upsilon}e^{\frac{iz^{2}}{4}}$,
which implies
\begin{equation}
{\psi_{11}}(z)=e^{\frac{\pi\upsilon}{4}}{D_a}\left(e^{\frac{\pi i}{4}}z\right)
\end{equation}
and
\begin{equation}
{\beta_{12}}{\psi_{22}}(z)={\beta_{12}}e^{-\frac{3\pi\upsilon}{4}}{D_{-a}}\left(e^{\frac{3\pi i}{4}}z\right).
\end{equation}
Consequently,
\begin{equation}
{\psi_{12}}(z)={\beta_{12}}e^{\frac{\pi\upsilon}{4}}e^{-\frac{3\pi i}{4}}{D_{-a-1}}\left(e^{-\frac{\pi i}{4}}z\right).
\end{equation}
Along the ray arg $z=-\frac{\pi}{4}$, ${\psi_+}(z)={\psi_-}(z)\left(
\begin{array}{cc}
0 & \overline{r({z_0})} \\
0 & 0 \\
 \end{array}
 \right)$ and
 ${\beta_{12}}e^{\frac{\pi(\upsilon-3i)}{4}}{D_{-a-1}}\left(e^{-\frac{\pi i}{4}}z\right)
 =e^{\frac{\pi\upsilon}{4}}{D_{a}}\left(e^{\frac{\pi i}{z}}\right)\overline{r({z_0})}
 +{\beta_{12}}e^{\frac{\pi(i-3\upsilon)}{4}}{D_{-a-1}}\left(e^{\frac{3\pi i}{4}}z\right)$. From Eq.(3.142), we have
\begin{equation}
{D_a}\left(e^{\frac{\pi i}{4}}z\right)=\frac{\Gamma(a+1)e^{\frac{i\pi a}{2}}}{\sqrt{2\pi}}{D_{-a-1}}\left(e^{\frac{3\pi i}{4}}z\right)
+\frac{\Gamma(a+1)e^{-\frac{i\pi a}{2}}}{\sqrt{2\pi}}{D_{-a-1}}\left(e^{-\frac{\pi i}{4}}z\right).
\end{equation}
Therefore, we obtain
\begin{equation}
{\beta_{12}}=\frac{e^{(\frac{3\pi i}{4}+\frac{\pi\upsilon}{2})}\Gamma(1+a)}{\sqrt{2\pi}}\overline{r({z_0})}
=\frac{e^{\left(\frac{-3\pi i}{4}+\frac{\pi\upsilon}{2}\right)}\upsilon\Gamma(i\upsilon)}{\sqrt{2\pi}}\overline{r({z_0})}.
\end{equation}
Finally, theorem (1.1) gives the long-time asymptotics of Eq.(1.8).
\section{Conclusion}  
In this paper, we study long-time asymptotics of complex mKdV equation by Defit-Zhou steepest descent method, in which the Schwartz initial data ${u_0}(x)\in \mathcal{S}(\mathbb{R})$ is inquired. The corresponding result is given in theorem 1.1.

In short order, we plan to consider weaker weighted Sobolev initial data ${u_0}(x)\in H^{{1,1}}(\mathbb{R})$ and ${u_0}(x)\in H^{{2,2}}(\mathbb{R})$ of long-time asymptotics to some NLPDEs.

\section*{Acknowledgements}

This work was supported by the National Natural Science Foundation of China (grant No.11971475).\\

\section*{Data Availability}
Data sharing is not applicable to this article as no new data were created or analyzed in this study.


\end{CJK*}
\end{document}